\documentclass{aa}  

\usepackage{graphicx}
\usepackage{subfigure}

\usepackage{natbib}
\usepackage[colorlinks, citecolor=blue, linkcolor=blue]{hyperref}
\usepackage{ulem} 

\usepackage{txfonts}

\usepackage{xcolor}

\begin{document}

   \title{J-PAS: A Neural Network Approach to Single Stellar Population Characterization}

   \titlerunning{NNs for Stellar Populations in J-PAS}

   \author{H. Domínguez Sánchez,
          \inst{1, 2}
          P. Coelho,
          \inst{4}
          G. Bruzual,\inst{5}
          A. Hernán-Caballero,\inst{2, 3}
          C. López Sanjuan,\inst{2, 3}
          J. A. Fernandez-Ontiveros,\inst{2, 3}
          L.A. Díaz-García,\inst{6}
          L. Suelves,\inst{7}
          A. Álvarez-Candal,\inst{6}
          I. Breda,\inst{6}
          S. Gurung-López,\inst{8, 9}
          V. Placco,\inst{10}
          J. Vega-Ferrero,\inst{2}
          J. M. Vílchez,\inst{6}
          R. Abramo,\inst{11}
          J. Alcaniz,\inst{12}
          N. Benitez,\inst{}
          S. Bonoli,\inst{13} 
          S. Carneiro,\inst{12}
          J. Cenarro,\inst{2, 3}
          D. Cristóbal-Hornillos,\inst{2}
          R. Dupke,\inst{12}
          A. Ederoclite,\inst{2, 3} 
          C. Hernández–Monteagudo,\inst{14, 15}
          A. Marín-Franch,\inst{2, 3} 
          C. Mendes de Oliveira,\inst{11}
          M. Moles,\inst{2}
          L. Sodré Jr,\inst{11}
          K. Taylor,\inst{16}
          J. Varela,\inst{2}
          \and
          H. Vázquez Ramió\inst{2, 3}
          }
    
    \authorrunning{Domínguez Sánchez et al.}      

   \institute{
   Instituto de Física de Cantabria, Av. de los Castros, 39005 Santander, Cantabria, Spain\\
              \email{helenad@ifca.es}
    \and         
   Centro de Estudios de Física del Cosmos de Aragón, Plaza de San Juan, 1, E-44001 Teruel, Spain 
   \and
  Unidad Asociada CEFCA-IAA, CEFCA, Unidad Asociada al CSIC por el IAA y el IFCA, Plaza San Juan 1, 44001 Teruel, Spain
     \and
   Universidade de São Paulo, Instituto de Astronomia, Geofïsica e Ciências Atmosféricas, Rua do Matão, 1226, 05508-090, São Paulo, SP, Brazil  
    \and
    Instituto de Radioastronom\'ia y Astrof\'isica, Universidad Nacional Aut\'onoma de M\'exico, Morelia, Michoac\'an 58089, M\'exico
      \and 
      Instituto de Astrofísica de Andalucía (CSIC), PO Box 3004, 18080, Granada, Spain
    \and Tartu Observatory, University of Tartu, Observatooriumi 1, Tõravere 61602, Estonia
    \and Observatori Astronòmic de la Universitat de València, Ed. Instituts d’Investigació, Parc Científic. C/ Catedrático José Beltrán, n2, 46980 Paterna, Valencia, Spain
    \and Departament d’Astronomia i Astrofísica, Universitat de València, 46100 Burjassot, Spain
    \and NSF NOIRLab, Tucson, AZ 85719, USA
       \and Departamento de Astronomia, Instituto de Astronomia, Geofísica e Ciências Atmosféricas, Universidade de São Paulo, São Paulo, Brazil
    \and Observatório Nacional, Rua General José Cristino, 77, São Cristóvão, 20921-400, Rio de Janeiro, RJ, Brazil
   \and Donostia International Physics Center (DIPC), Manuel Lardizabal Ibilbidea, 4, San Sebastián, Spain
   \and  Instituto de Astrofísica de Canarias, C/ Vía Láctea, s/n, E-38205, La Laguna, Tenerife, Spain 
   \and  Universidad de La Laguna, Avda Francisco Sánchez, E-38206, San Cristóbal de La Laguna, Tenerife, Spain
   \and  Instruments4, 4121 Pembury Place, La Canada Flintridge, CA 91011, U.S.A.
     }


 
  \abstract
   {J-PAS (Javalambre Physics of the Accelerating Universe Astrophysical Survey) will present a groundbreaking photometric  survey covering 8500 deg$^2$ of the visible sky from Javalambre, capturing data in 56 narrow band filters. This survey promises to revolutionize galaxy evolution studies by observing $\sim$10$^8$ galaxies with low spectral resolution. A crucial aspect of this analysis involves predicting stellar population parameters from the observed galaxy photometry.  In this study, we combine the exquisite J-PAS photometry with state-of-the-art single stellar population (SSP) libraries to accurately predict stellar age, metallicity, and dust attenuation with a neural network (NN) model. The NN is trained on synthetic J-PAS photometry from different SSP librares  (E-MILES, Charlot \& Bruzual, XSL), to enhance the robustness of our predictions against individual SSP model variations and limitations. To create mock samples with varying observed magnitudes we add artificial noise in the form of random Gaussian variations within typical observational uncertainties in each band. Our results indicate that the NN can accurately estimate stellar parameters for SSP models without evident degeneracies, surpassing a bayesian SED-fitting method on the same test set. We obtain median bias, scatter and percentage of outliers $\mu$ = (0.01~dex, 0.00~dex, 0.00~mag), $\sigma_{NMAD}$ = (0.23 dex, 0.29 dex, 0.04 mag), f$_{o}$ = (17\%, 24\%, 1\%)  at $ i \sim$17 mag for age, metallicity and dust attenuation, respectively. The accuracy of the predictions is highly dependent on the signal-to-noise (S/N) ratio of the photometry, achieving robust predictions up to $i$ $\sim$ 20 mag. 
}

   \keywords{stellar populations --
                neural networks --
                photometry
               }

\maketitle


   \begin{figure*}
   \centering
   \includegraphics[width=0.3\hsize]{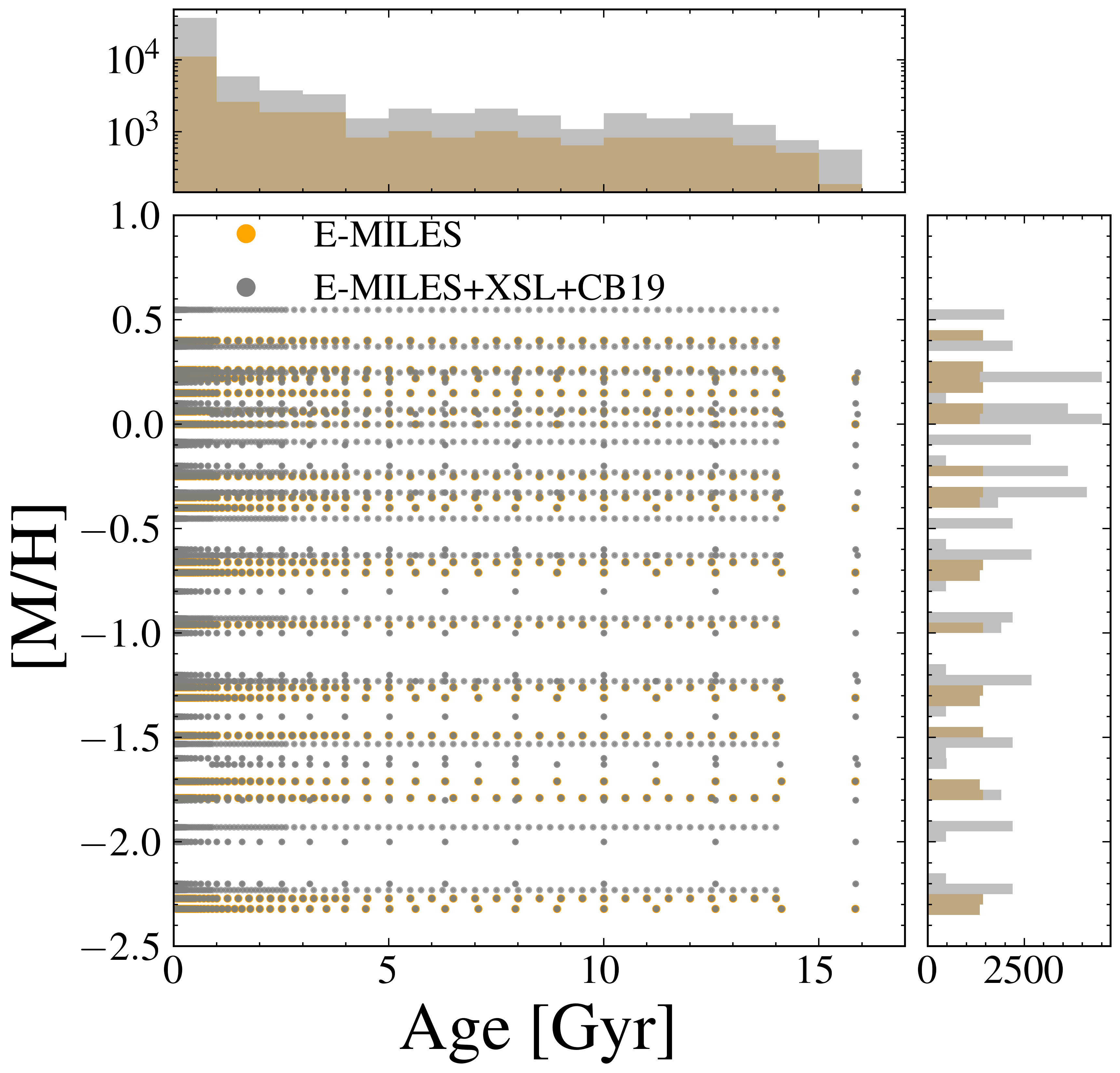}
    \includegraphics[width=0.3\hsize]{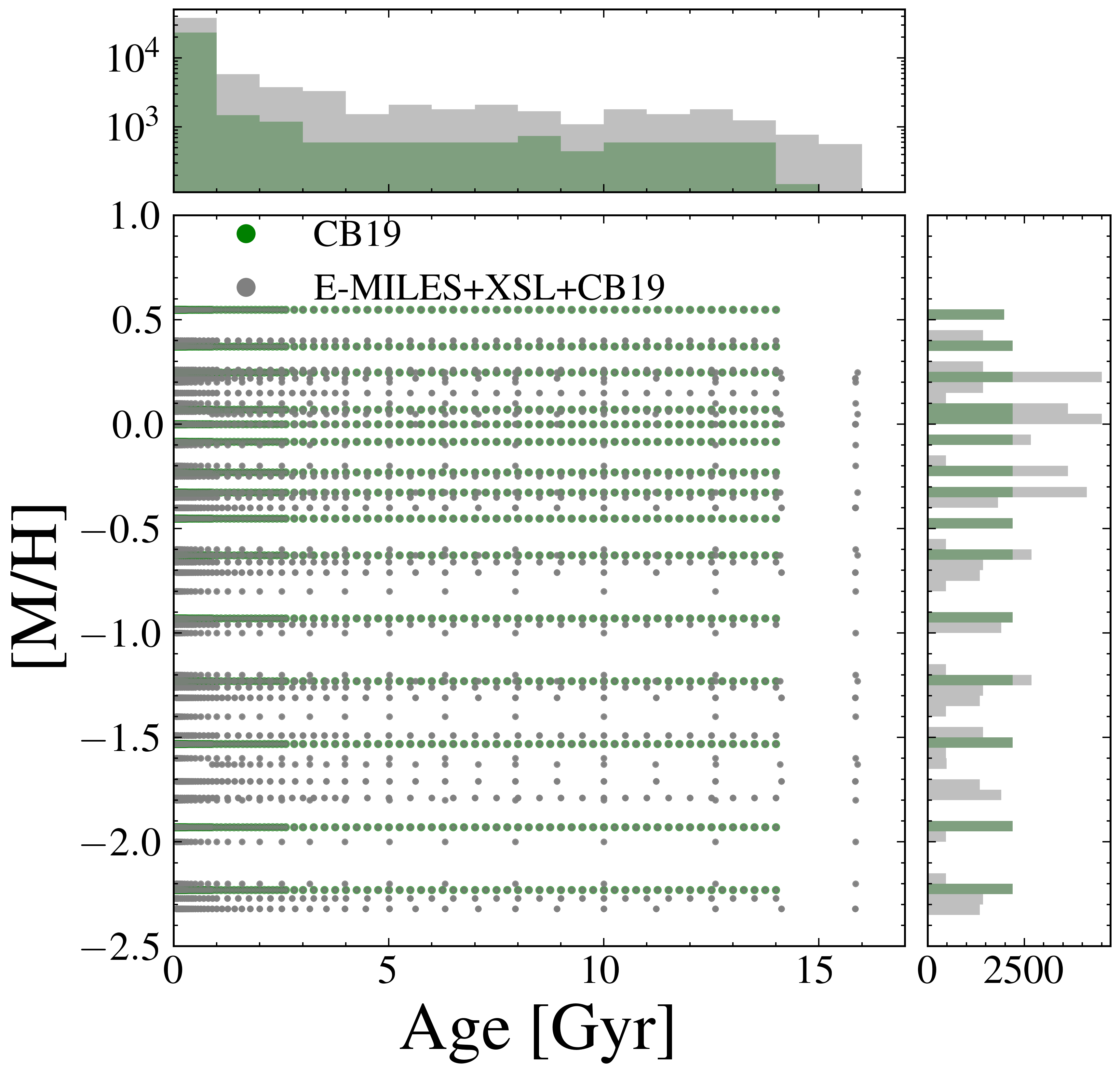}
    \includegraphics[width=0.3\hsize]{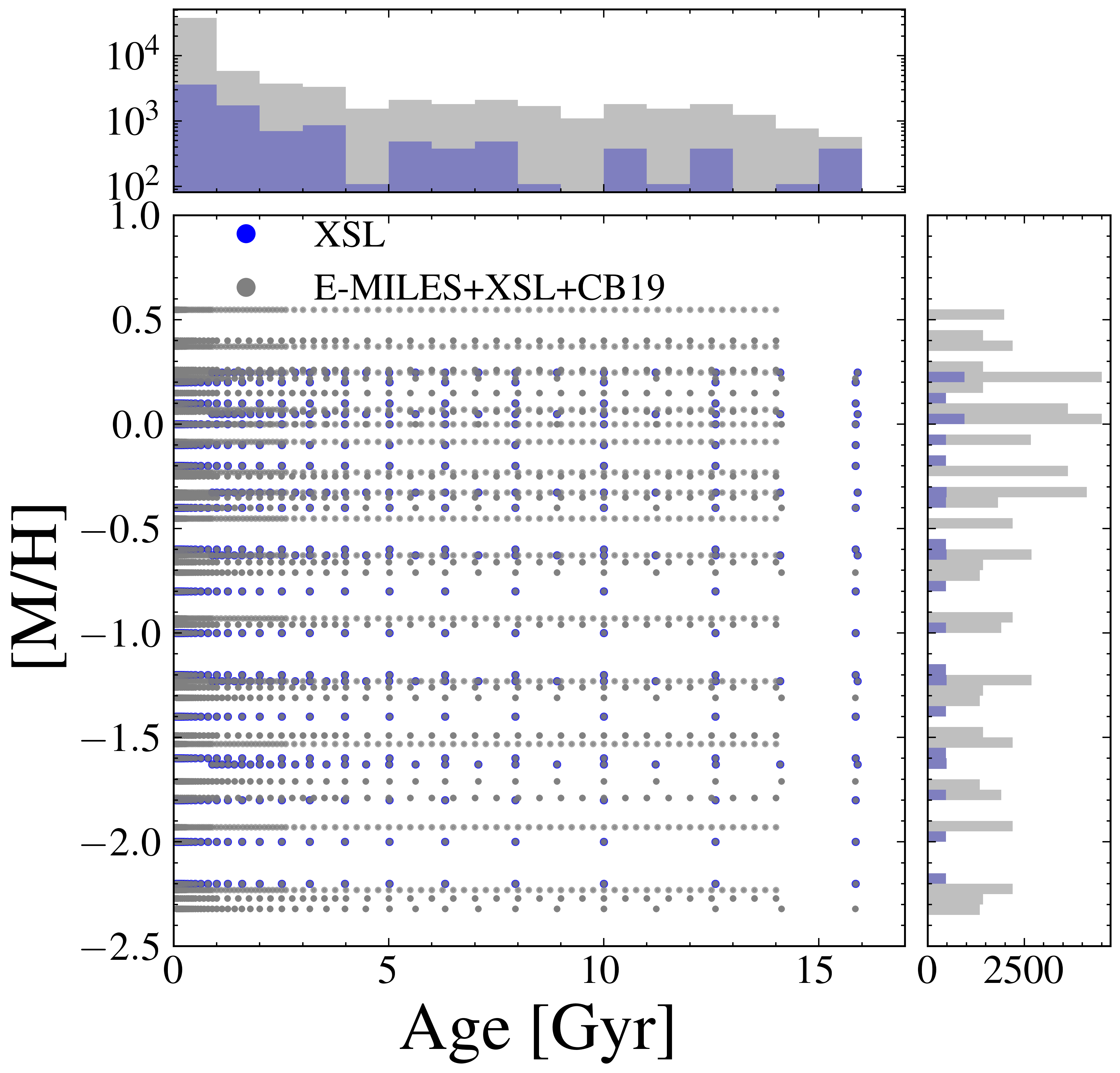}
   \caption{Age-metallicity coverage of the three SSP synthesis models used in this work to train the NNs. The grey symbols (dots and histograms) show the combination of the three SSP libraries, while in each panel we show the individual libraries: E-MILES in orange (left panel), CB19 in green (middle panel) and XSL in blue (right panel). Note that the models with ages below < 30 Myr are only included in the  CB19 SSP library.}
              \label{fig:SSPgrid}%
    \end{figure*}

\section{Introduction}

The field of galaxy evolution aims at understanding how the properties of galaxies have changed across cosmic time and how these properties relate with each other. The main attributes of galaxies are the morphology,  the environment and the stellar populations, i.e., the stars contained within them. The assembly histories of galaxies leave their imprint in the stellar populations and interstellar medium, that can be characterised by the total stellar mass, the mean age of the stellar population, the metallicity and the dust content, among other parameters. Unveiling the stellar population of large numbers of galaxies is fundamental to gain knowledge on the processes that gave rise to  the great variety of galaxies that we see in the Universe. 

Single stellar population (SSP) models combine spectral stellar libraries with isochrones to predict the time-evolution of the light emitted by a galaxy with a given Initial Mass Function (IMF) and metallicity [M/H]. By comparing these theoretical models with the observations, the stellar properties of the galaxies can be estimated. This can be done through full-spectral fitting \citep[e.g.][]{cid+05, Cappellari2011} or spectral index measurements \citep[e.g.][]{trager+00a, serven+05, Eftekhari+21} when spectroscopic observations or narrow-band photometry are available (e.g., \citealt{HernanCaballero2013, HernanCaballero2014, DS2016}), or by means of Spectral Energy Distribution fitting (SED-fitting, \citep[e.g.][]{Boquien2019, DiazGarcia2015,  prospect20}, for photometric observations. 

The better spectral resolution of spectroscopy allows more accurate measurements of the stellar populations, but spectroscopic observations are very time consuming and often require pre-selected targets, biasing the analysis to relatively bright objects and smaller samples. On the other hand, photometric surveys are only limited by their depth and usually cover large areas of the sky, but their insufficient spectral resolution complicates the estimation of stellar populations, sharpening the well know age-metallicity-attenuation degeneracies (e.g., \citealt{DiazGarcia2015}). In this work we take advantage of the Javalambre Physics of the Accelerating Universe Astrophysical Survey (J-PAS), providing photometry in 56 narrow bands (plus the i-band for detection) that provides pseudo-spectra with a resolution R $\sim$ 60 and it is almost complete up to $i$ < 22.0 mag. J-PAS will observe roughly 1/5 of the sky, providing extremely valuable information for hundreds of millions of galaxies \citep{Benitez2014, Bonoli2021}. Extracting stellar population parameters for these wealthy dataset is a challenging and time consuming task, but the reward is worth it, since the scientific applications of these estimates are innumerable.

  While traditional SED-fitting codes have been widely utilized for this purpose, they often require a predefined set of templates for comparison, which can be computationally intensive. Moreover, a limited parameter space may fail to adequately represent the galaxy population. In recent years, there has been a surge in the application of machine learning algorithms, particularly neural networks (NNs), for various tasks in astronomy (e.g., \citealt{Whitten2019, Galarza2022, Martinez-Solaeche2022, Wang2022, Martinez-Solaeche2023, delPino2024, Gurung2022, Gurung2025}), particularly in image analysis (e.g.,\citealt{DS2018, DS2019a, Walmsley2022,  DS2023, Bom2024} to mention a few, see \citealt{Huertas-Company2023} for a complete review). However, their potential for predicting stellar populations of galaxies has remained barely unexplored \citep{Liew-Cain2021, Woo2024, Wang2024, IglesiasNavarro2024} offering considerable room for exploration.

In this work, we combine the high precision J-PAS photometry with cutting-edge SSP models (E-MILES, Charlot \& Bruzual, XSL)  to estimate stellar population parameters with NNs. We use synthetic photometry with realistic noise to train and test the NN models. In Sect. \ref{sect:data} we describe the SSP libraries and how the synthetic photometry is computed, Sect. \ref{sect:NN} gives details on the NN and training strategy. We show our results on the synthetic photometry in Sect.  \ref{sect:results} and compare them with previous works and SED-fitting results in Sect. \ref{sect:discussion}, while Sect. \ref{sect:JPAS} shows the predictions for a sample of real local galaxies observed by  the first J-PAS data release. Finally, Sect. \ref{sect:conclusions} summarises our findings.

\section{Data}
\label{sect:data}

In this section we describe the SSP libraries used in this work and how we generate the synthetic J-PAS photometry to train and test our NN for estimating SSP properties. 


   \begin{figure}
   \centering
   
 \includegraphics[width=0.99\hsize]{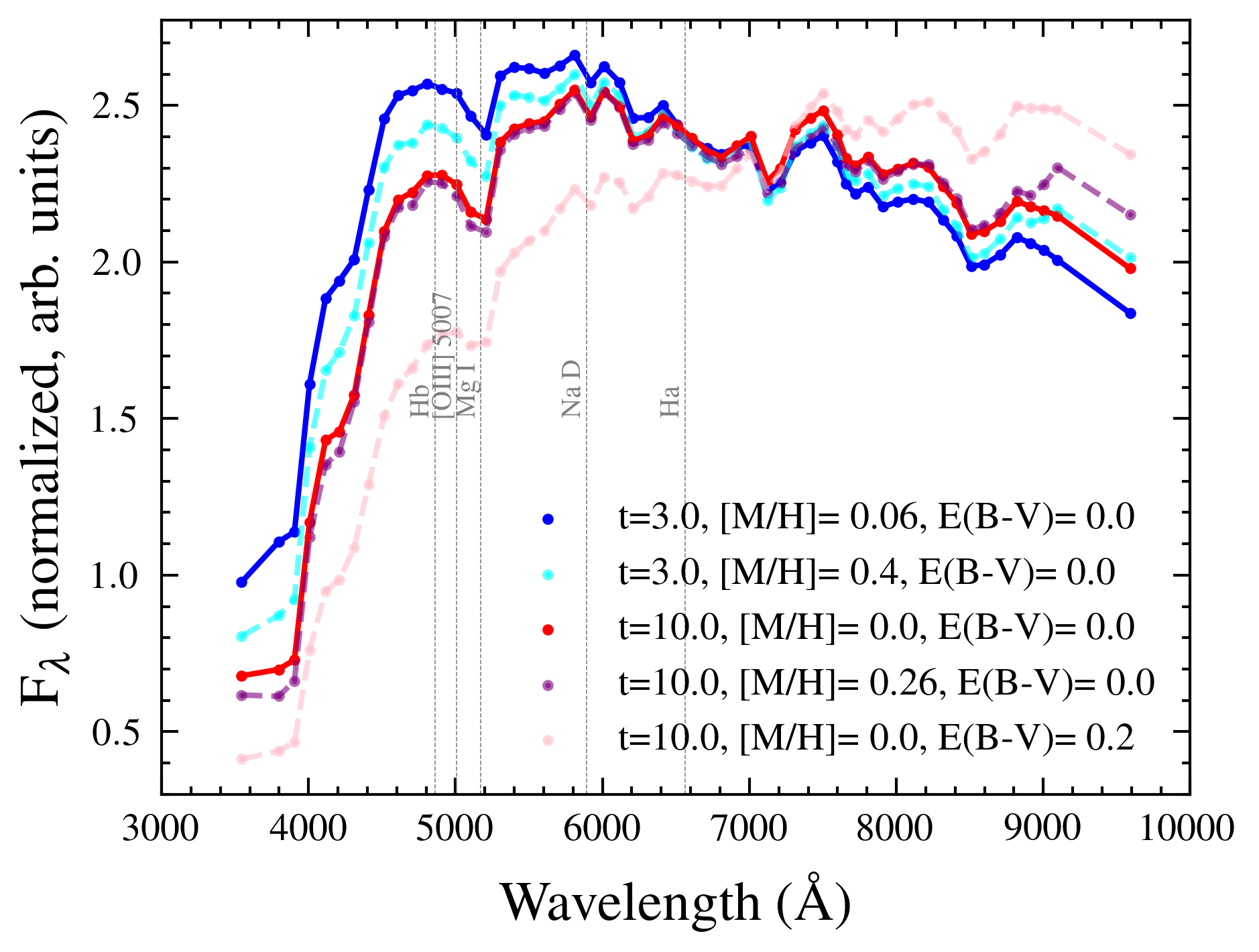}
   \caption{Examples of synthetic J-PAS photometry from E-MILES SSP for a young (3 Gyr) and an old (10 Gyr) SSP  with different metallicities and dust attenuation, as stated in the legend. The fluxes are normalized to the mean value of the flux for each SSP. Note the subtle differences that need to be distinguished for a proper characterization of the SSPs (e.g., compare the red and purple lines).}
    \label{fig:SED}%
    \end{figure}

       \begin{figure*}
   \centering
    \includegraphics[width=0.45\hsize]{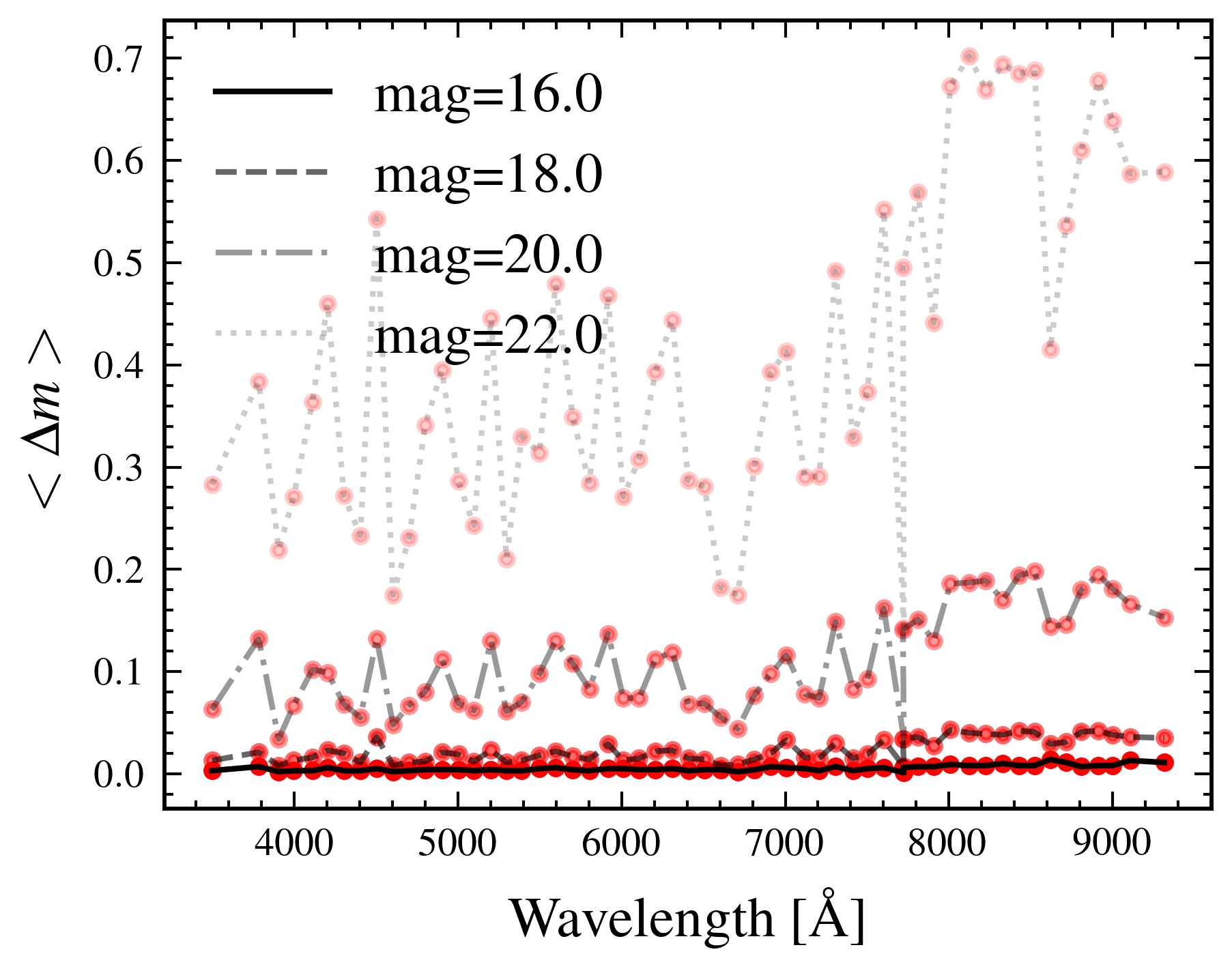}
   \includegraphics[width=0.45\hsize]{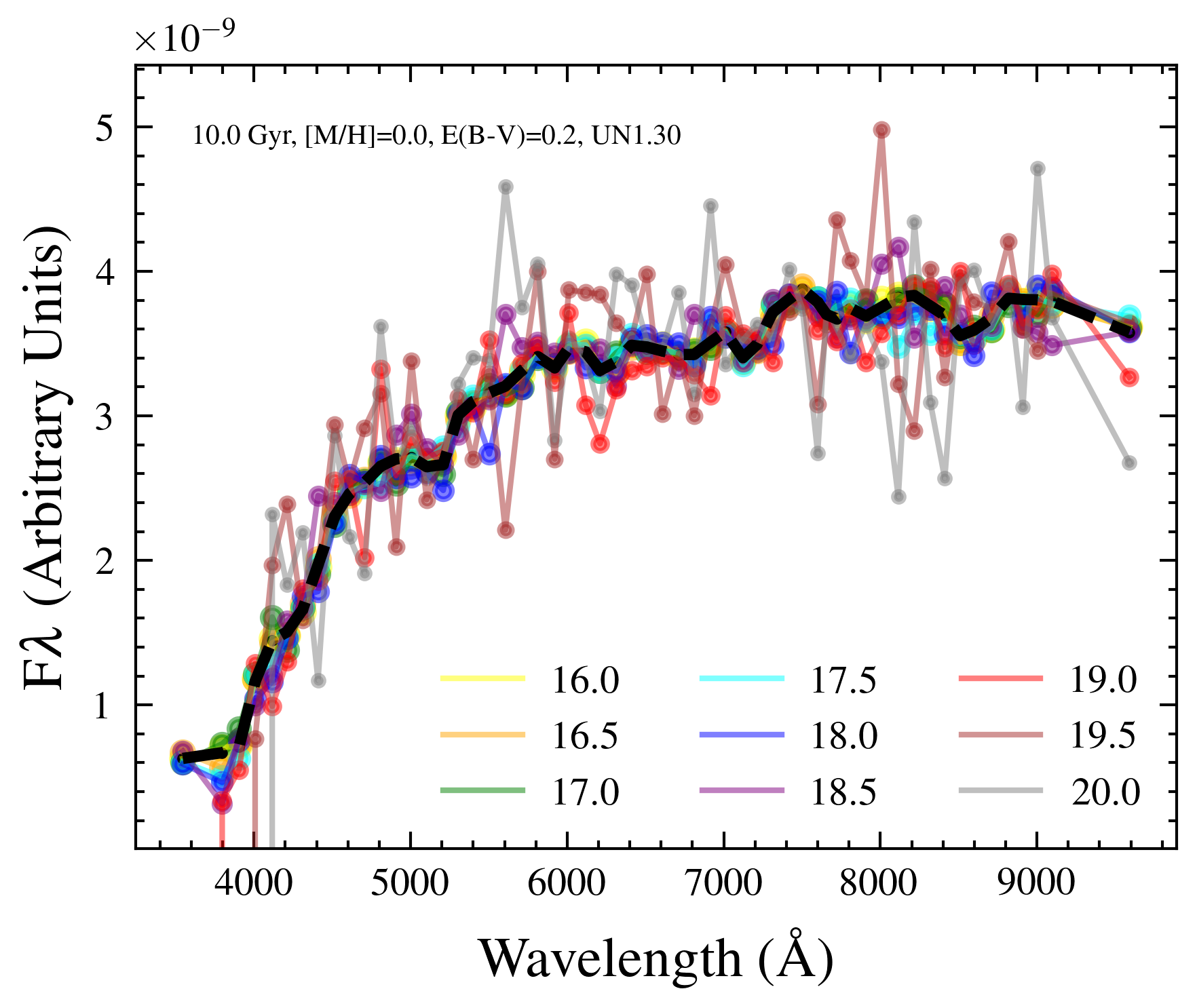}
   \caption{Left: average magnitude error of mini-JPAS galaxies for the 57 filters for different observed magnitudes (16, 18, 20 and 22 mag). Right: Synthetic J-PAS photometry obtained after applying random Gaussian variations consistent with the typical errors at different magnitudes. Colors represent different `mock' observed magnitudes. The thick dashed black line is the original SED.}
    \label{fig:SED-GV}%
    \end{figure*}

\subsection{Single stellar population models}

 In this work, we use supervised learning to train a NN to predict SSP properties (age, metallicity, dust attenuation) from J-PAS photometry. To ensure a reliable ground truth, we use SSP synthesis models, for which these parameters are well known, instead of real galaxies, where the stellar populations have to be derived by other means (e.g., SED or spectral fitting), with their associated uncertainties and biases.
 
 A SSP model is defined by a single age, metalicity and IMF, while galaxies are composed by mixtures of populations. Yet, when we refer to the "age" of a stellar population, the concept can be ambiguous, as its definition varies across different studies. In SED or spectral fitting, the most commonly used definitions are mass-weighted or light-weighted ages, but other definitions can be adopted (see Sect. 3.3 in \citealt{goncalves+20}). These codes adopt either non-parametric, like  \textsc{Muffit} \citep{DiazGarcia2015}, \textsc{Starlight} \citep{CidFernandes2011}, \textsc{pPXF}\citep{ppxf04}, or parametric star-formation histories, such as \textsc{BaySeAGal} \citep{GonzalezDelgado2021}, \textsc{ProSpect} \citep{Johnson2021}, \textsc{Cigale} \citep{Boquien2019}. In the non-parametric case, mixtures of an arbitrary number of SSPs require strategies to explore the parameter space in search of the minima, as it is not feasible to forward model all possible combinations of SSPs. In the case of parametric codes, the parametric form of the star formation history (SFH) and age-metallicity relations need to be adopted a priori, which often introduces biases and results in less accurate inferred parameters (e.g. \citealt{leja+19}). Flexible SFHs are being developed that approach the behavior of the non-parametric descriptions (e.g. \citealt{Bellstedt+20}), yet the forward modeling of galaxy SEDs from all possible combinations requires dedicated work.

In our analysis, we take a simpler approach and approximate the galaxy SED to one SSP, as was commonly adopted in classical stellar population work using spectral indices \citep[e.g.][]{trager+00a, thomas+05, vazdekis+10}. For instantaneous SFHs, the SSP-equivalent, light-weighted and mass-weighted parameters are equivalent. For extended SFHs, the correlation between the parameters has been investigated in the literature (e.g. \citealt{serra_trager07, trager_somerville09}). By analysisng a series of mock galaxies, these works show that SSP-equivalent and light-weighted ages correlate, with the light-weighted ages falling in between the SSP-equivalent and the mass-weighted ages. The SSP-equivalent age mostly measures the most recent star formation and can be interpreted as a lower limit of the ages in the studied population. On the other hand, the SSP-equivalent metallicities depend mainly on the chemical composition of the old population and closely track the mass- and light-weighted metallicities. For the goals of this paper and to keep our discussion more simple and clear, we think that limiting our analysis to SSPs is justified. We discuss the estimates of our NN -  trained on SSP - on composite stellar populations (CSP) in Appendix \ref{Appendix:CSP}.

We use three state-of-the-art SSP libraries, commonly used in the literature: E-MILES \citep{Vazdekis2016}, Charlot\&Bruzual 2019 \citep[CB19 hereafter]{Plat2019} and the X-shooter Spectral Library \citep[XSL]{Verro2022}. By combining different SSP libraries in our training we aim to make the model more robust against spectral resolution or any bias that may arise from the parameter space sampling of the SSP libraries. In addition, we note that, after several tests, we found a better performance of the NN predictions when the three SSP libraries were combined together in the training. Below we summarise some important characteristics of each SSP library. We refer the reader to the corresponding references for a more detailed description.


   \begin{figure*}
   \centering
   \includegraphics[width=0.33\hsize]{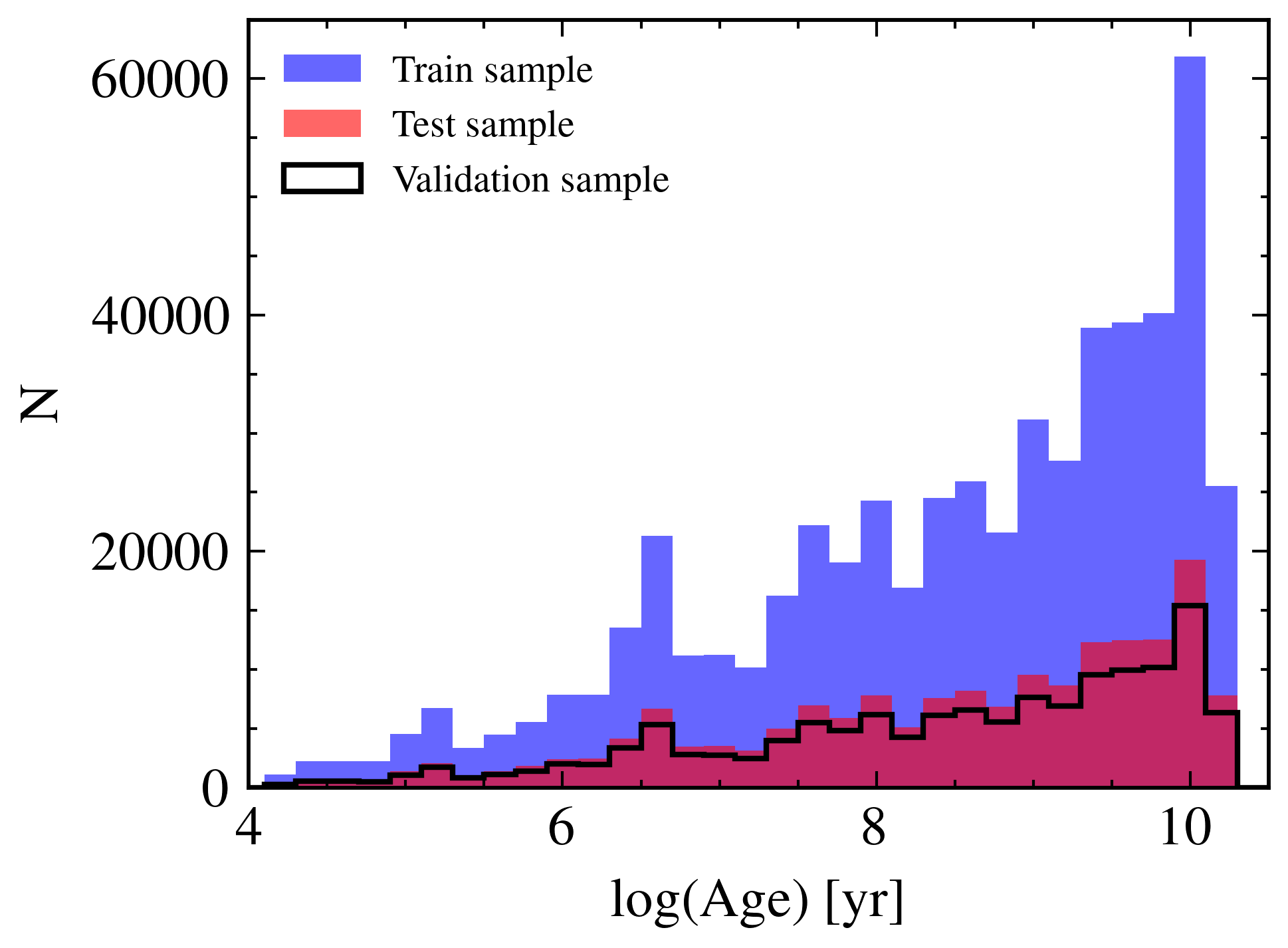}
    \includegraphics[width=0.33\hsize]{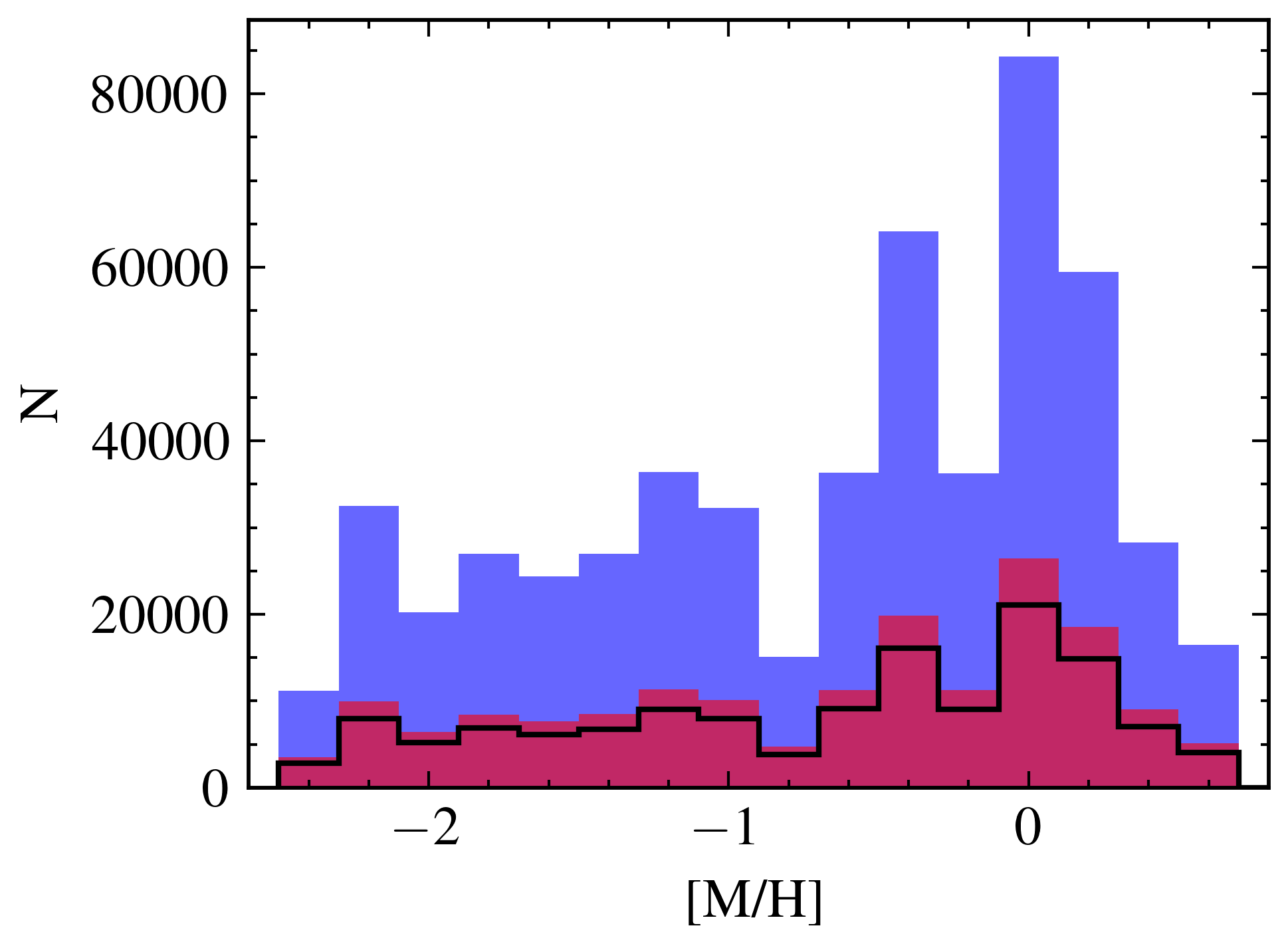}
    \includegraphics[width=0.33\hsize]{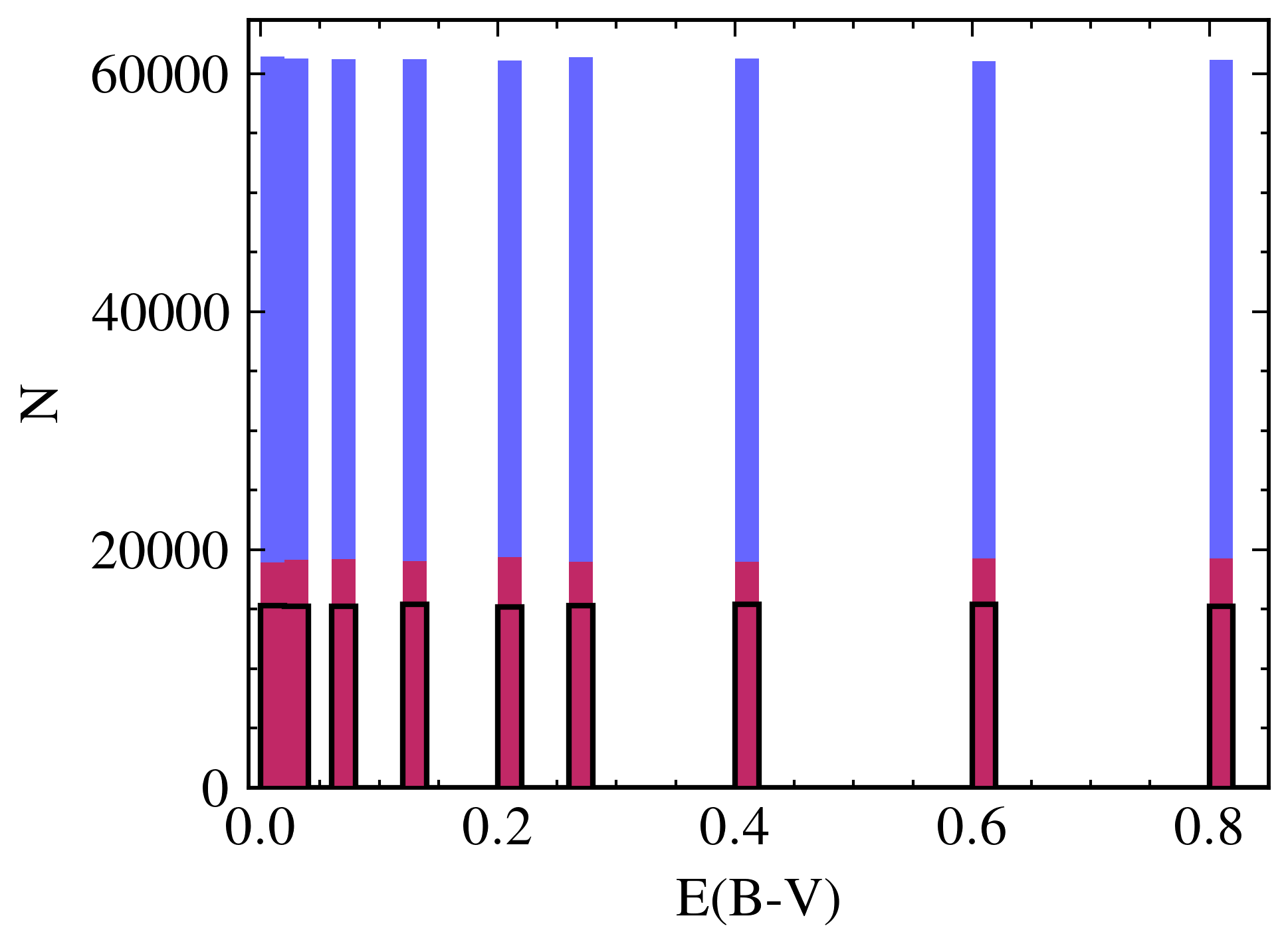} 
   \caption{Age (left panel), metallicity (middle panel) and dust attenuation  (right panel) distribution for the trainning (blue), test (red) and validation (empty black) sub-samples. The sub-samples are randomly selected, showing no differences in the distributions.}
    \label{Train-Test}
    \end{figure*}

\subsubsection{E-MILES library}
\label{sec:e-miles}
The UV-extended E-MILES stellar population synthesis models cover the spectral range  1680--50 000 \AA~at moderately high resolution. The  space-based New Generation Stellar Library  is used to compute spectra of single-age, single-metallicity stellar populations in the wavelength range from 1680 to 3540 \AA, joined with those  computed in the visible using MILES, and other empirical libraries for redder wavelengths. The isochrones used are Padova2000 and BaSTI \citep{Girardi2000, Pietrinferni2004}. The models span the metallicity range -2.32 $\le$ [M/H] $\le$ +0.4  and ages above 30 Myr (i.e., they do not include starburst and post-starburst populations), for a suite of initial mass function (IMF) types with varying slopes. In particular, in this work, we use the \cite{Chabrier2003}, \cite{Kroupa2001} and Unimodal $\Gamma=1.3$ IMFs. The original models have no dust attenuation. We have generated a new set of models by reddening each of the original models to  different values\footnote{The E(B-V) values used in this work are 0, 0.03, 0.06, 0.13, 0.26, 0.40, 0.60 and 0.80.} using the 
law from \citet{cardelli+89} with $R = 3.1$. The left panel of Fig. \ref{fig:SSPgrid} shows the age-metallicity parameter space covered by the 26622 E-MILES models used in this work compared to the parameter spaced covered by the three SP libraries combined together.

\subsubsection{Charlot \& Bruzual 2019 library}

The CB19 stellar population synthesis models cover the spectral range  5-80 000 \AA~at moderate-to-high spectral resolution  using stellar fluxes from the literature, PARSEC isochrones and \cite{Chabrier2003} IMF. The age-metallicty coverage of the  29700 CB19 models used in this work is shown in the middle panel of Fig. \ref{fig:SSPgrid}, covering -2.23 $\le$ [M/H] $\le$ +0.54  and ages above 1.5 Myr, being the library including the youngest populations. As explained for the E-MILES, we have generated a new set of models with different E(B-V) values.


\subsubsection{XSL library}

The XSL SSP models are based on the empirical X-shooter Spectral Library DR3 \citep{Verro2022-lib}, a moderate-to-high resolution, near-ultraviolet-to-near-infrared (3500--24800 \AA, R\,$\sim$\,10\,000) spectral library, composed of 830 stellar spectra of 683 stars.  The wavelength coverage and careful modelling of the RGB and AGB fluxes is optimised to bridge optical and NIR studies of intermediate-age and old stellar populations. The models span the metallicity range --2.2 $<$ [Fe/H] $<$ +0.25 and ages above 50 Myr (i.e., no starburs or post-starburst are included), with PARSEC and Padova2000 isochrones and two IMF parametrizations (\citealt{Salpeter1955} and \citealt{Kroupa2001}).  We have also generated a new set of models with different E(B-V) values for this library. The age-metallicity covered by the 9936 XSL models used in this work is shown in the righ panel of Fig. \ref{fig:SSPgrid}.

   \begin{table}
      \caption[]{Number of SSP models for the different libraries for the original sample and the `mock' samples with different S/N. }
         \label{tab:Nmodels}
     $$ 
         \begin{array}{p{0.5\linewidth}ll}
            \hline
            \noalign{\smallskip}
            SSP model     &  N  &  N_{mock}\\
            \noalign{\smallskip}
            \hline
            \noalign{\smallskip}
                E-MILES &  26622 & 346086  \\
            BC19   &  29700 &  386100  \\
            XSL    &  9 936   & 129168  \\
            \noalign{\smallskip}
            \hline
         \end{array}
     $$ 
   \end{table}

\subsection{J-PAS photometry}

J-PAS is a wide field Cosmological Survey performed with the 1.2 Gigapixel JPCam instrument of the 2.5 m JAST250 telescope at the Javalambre Observatory (Teruel, Spain). J-PAS observations, started in October 2023, will eventually cover $\sim$8500 deg$^2$ of the northern sky \citep{Benitez2014,Bonoli2021}. Its unique set of 56 narrow bands (FWHM $\sim$ 140 \AA) spanning the optical range from 3500 \AA{} to 9300 \AA{} allows excellent photometric redshift accuracy \citep[$\sigma_z$=0.003;][]{HernanCaballero2021} and precise measurement of the stellar populations of galaxies (e.g., \citealt{Benitez2014, MejiaNarvaez2017, GonzalezDelgado2021}).


The synthetic J-PAS photometry is generated by convolving the SSP spectra described in Section \ref{sect:data} with the official J-PAS filter transmission curves\footnote{Obtained from \url{http://svo2.cab.inta-csic.es/theory/fps/}, see \citet{Rodrigo2024}}. Dust attenuation is applied directly to the spectra before computing the photometry (see details in Section \ref{sec:e-miles}). 
Fig. \ref{fig:SED} shows examples of the synthetic photometry for a young and an old SSP with different metallicities and dust attenuation from the E-MILES library. Some of the spectral features are clearly visible in the synthetic fluxes (e.g., Mg and Na absorption). Note that the models are at rest-frame wavelength (z=0) and have no noise. Since the objective of this project is to predict SP from observed galaxies, we need to model and apply noise to the synthetic photometry to make the data more similar to real observations. In order to do so, we apply random Gaussian variations to the SSP fluxes, where the amplitude $\sigma$ of the variations is derived as the average of the photometric error of the galaxies in mini-JPAS\footnote{Mini-JPAS is a 1 deg$^2$ survey of the AEGIS field observed with 56 J-PAS filters on the Pathfinder camera. More details in \cite{Bonoli2021}}, $\Delta m$ \footnote{We transform magnitude errors in signal to noise following the formula  S/N = 2.5/ln(10)$\Delta m$.}, as a function of AUTO magnitude in each band. Fig. \ref{fig:SED-GV} shows these errors (in magnitude) for each band at different observed magnitudes. We generate synthetic spectra of `mock' observed magnitudes from $i$ = 16.0 - 22.0 mag in 0.5 magnitude bins, thus multiplying by 13 the number of SSP models. As an example, Fig. \ref{fig:SED-GV} illustrates one of these `mock' SSPs with different levels of noise (corresponding to different `observed' magnitudes). The number of SSP models from each library (original and with different levels of noise) are reported in Table \ref{tab:Nmodels}.

   \begin{figure*}
   \centering
   \includegraphics[width=0.49\hsize]{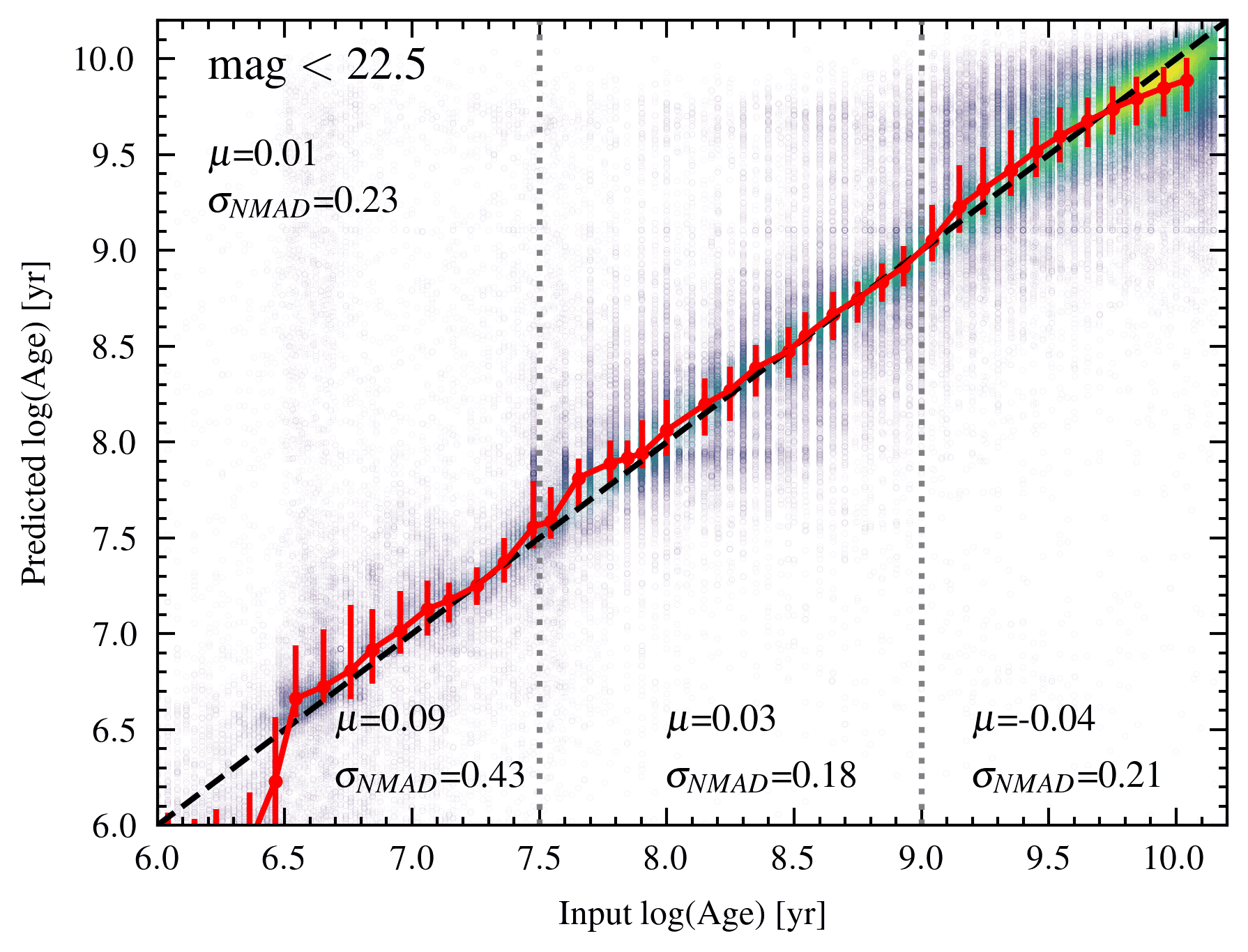}
    \includegraphics[width=0.49\hsize]{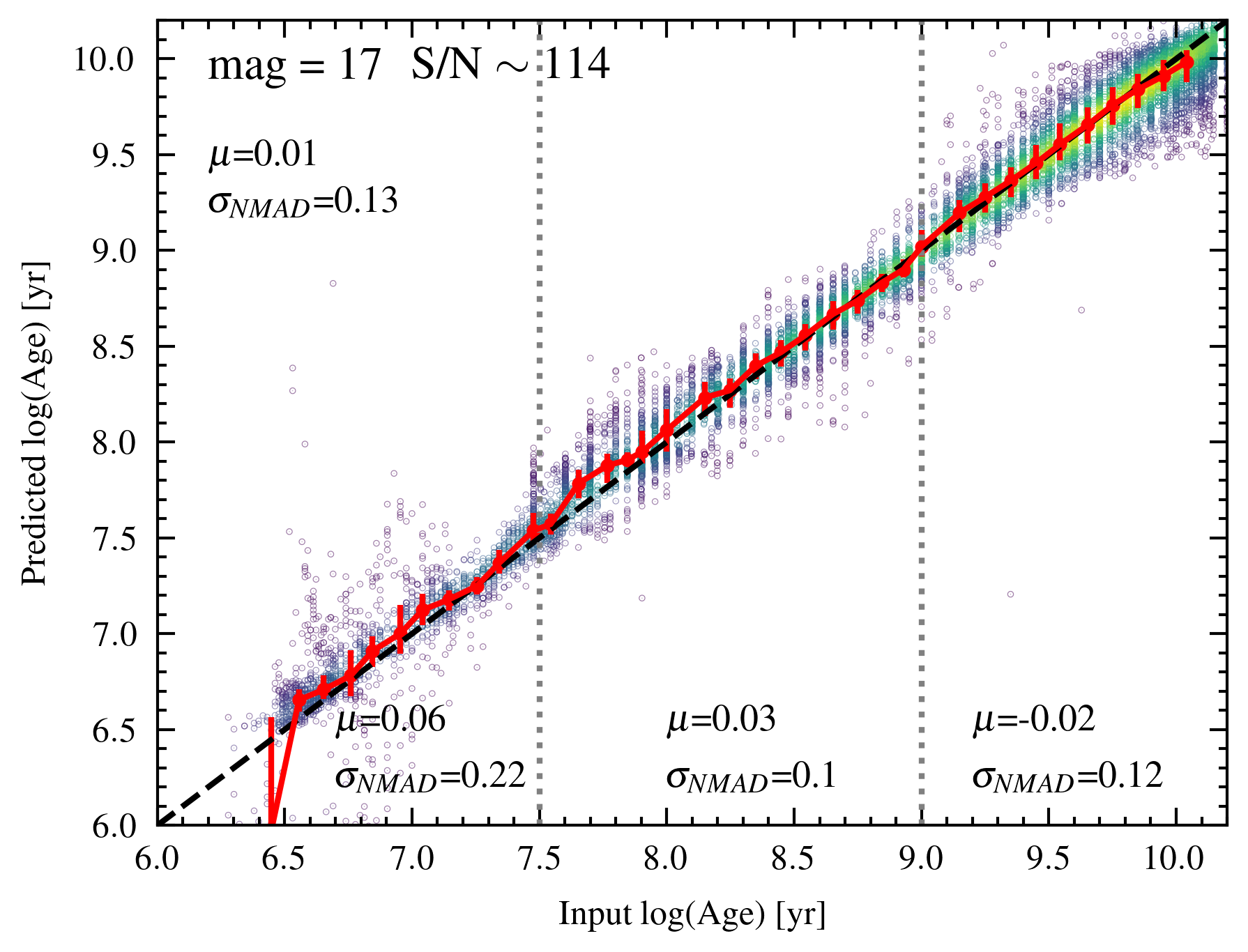}
    \includegraphics[width=0.49\hsize]{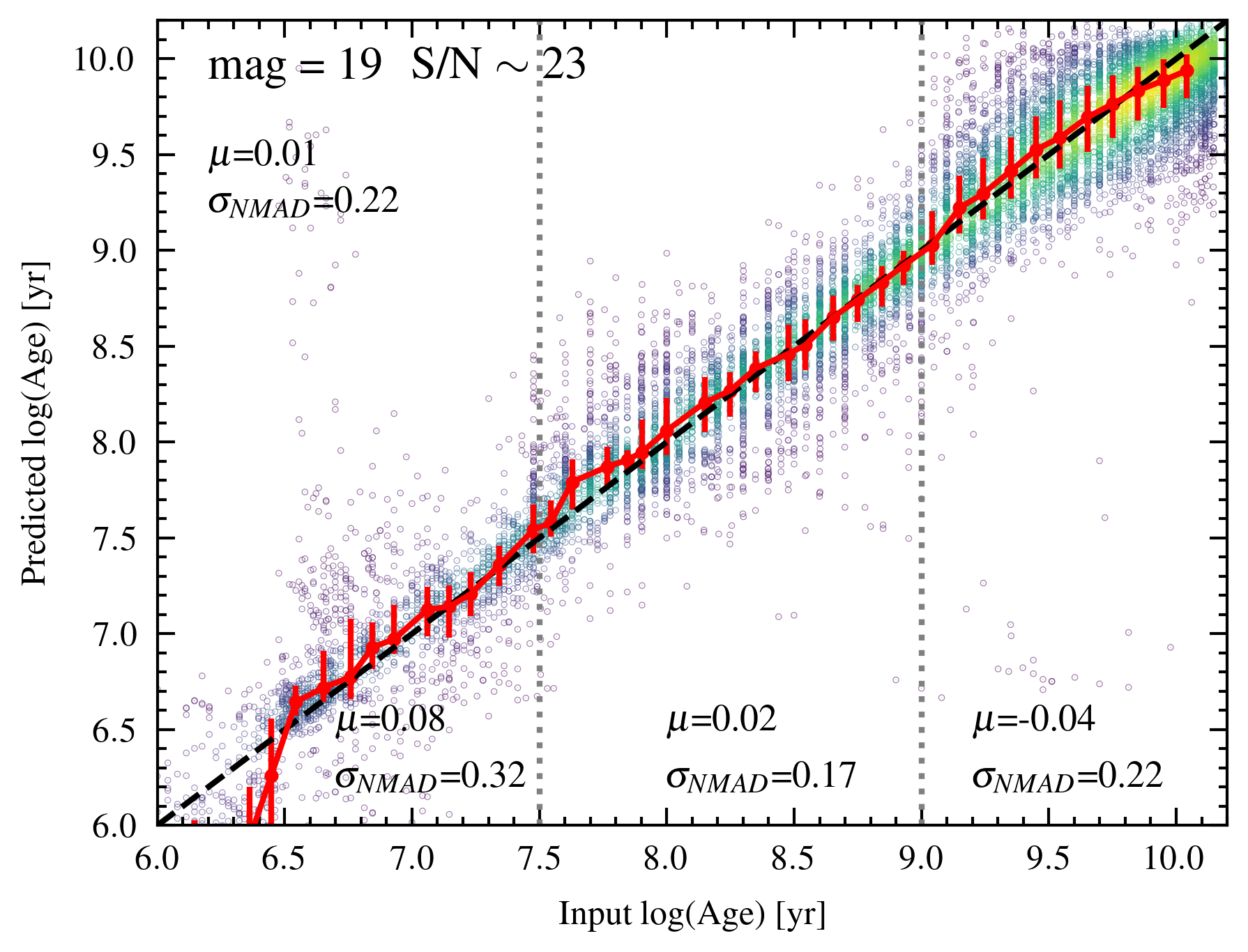} 
    \includegraphics[width=0.49\hsize]{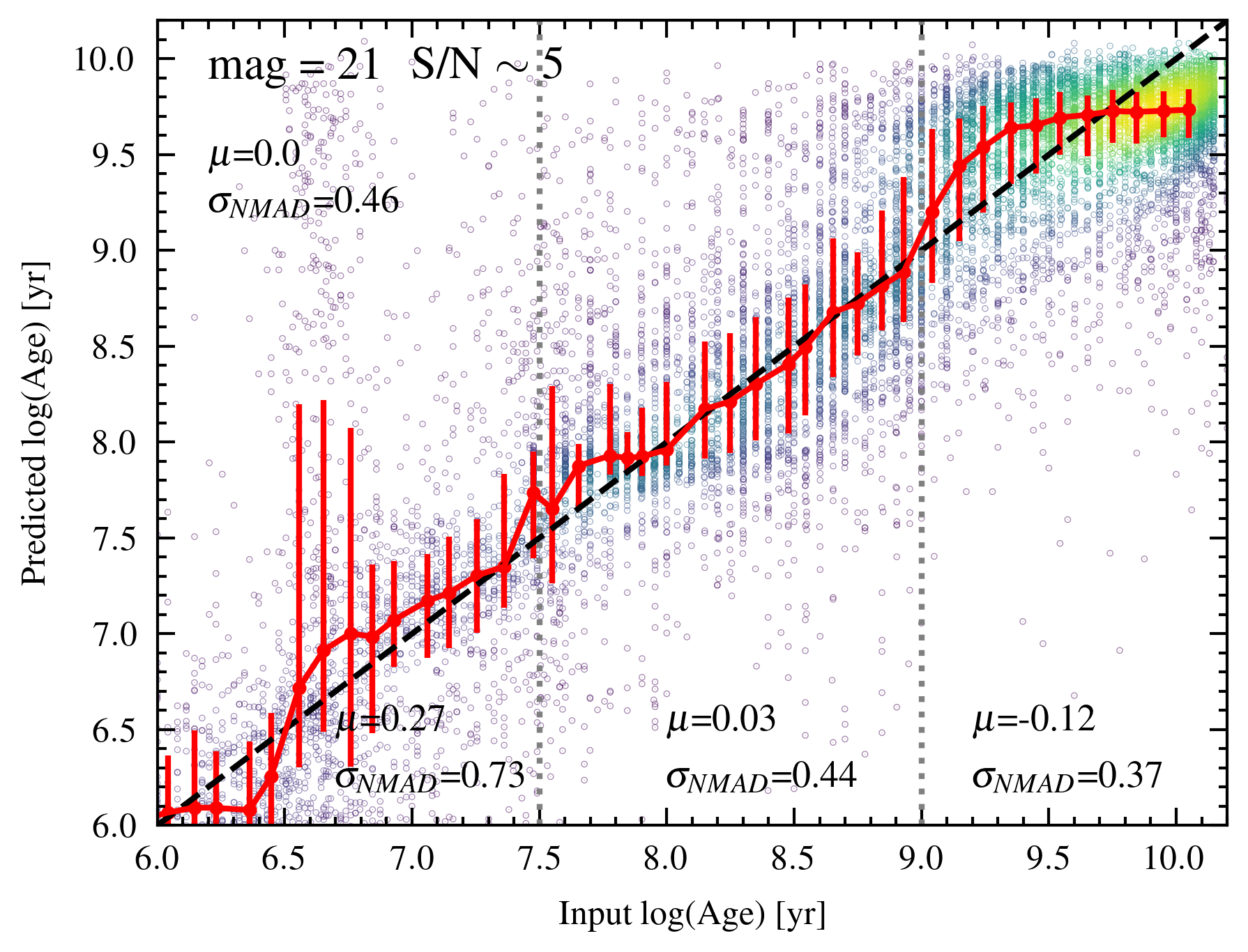} 
   \caption{Predicted age vs input age in different magnitude bins. The  median $\mu$ and $\sigma_{NMAD}$ of the residuals ($\Delta$ = output - input) are reported for the full sample, and for the three age ranges delimited by the dashed vertical lines.  Red points indicate the median values in age bins, with red error bars showing the interquartile range (25th to 75th percentiles) of the distribution. Symbols are color-coded by number density (more populated regions are plotted in yellow). The upper left panel shows the results for all the test sample, while the other three panels are limited to a given $i$ magnitude bin, stated in the legend.  The typical S/N  in the narrow-band filters is also reported.}
    \label{Age}%
    \end{figure*}

\section{Neural Network: architecture and training strategy}   
\label{sect:NN}

We employ a simple neural network  (NN) architecture to train our  machine learning model, taking as input a data vector of 57 dimensions with the J-PAS flux in each filter (in F$_{\nu}$, normalised by the median value of the flux of each SSP model). The labels correspond to the SP parameters (age\footnote{In yr in logarithmic scale}, [M/H] and E(B-V)). We explored two different  NN configurations:
\begin{itemize}
\item base NN: a hidden layer with 128 neurons, three  hidden layers with 256 neurons each and a  dense layer returning the output (172289 free parameters)
\item SED NN\footnote{Following \cite{Kaeufer2023}}: a hidden layer with 128 neurons, six  hidden layers with 300 neurons each and a  dense layer returning the output (497925 free parameters)
\end{itemize}
We find better results with the base NN configuration, despite its smaller number of free parameters. This suggests that a simpler model is suitable for the analysis presented in this work, preventing overfitting.  We train for 100 epochs, using mean square error (\textsc{MSE}) as loss function, \textsc{adam} optimizer and learning rate $lr=0.002$. Better results are obtained when using $batch~size = 32$ rather than 100 or 1000, even when combined with dropout. In particular, a small batch size significantly reduced overfitting. The comparison is shown in Fig. \ref{A-Metrics} in Appendix \ref{Appendix:models}. We train three independent NNs (i.e., the output dimension is one), one for each of the parameters (age, metallicity, attenuation). We found better results with this configuration rather than  training a single model with a three-dimensional output.

We randomly split our sample into train/validation/test subsamples with a  65/15/20 proportion, resulting in 551253, 172263 and 137838 SSP models, respectively. The distributions of the input SSP parameters for each subset are shown in Fig. \ref{Train-Test}. The loss plots showing the training behavior are reported in Appendix  \ref{Appendix:Loss}.

\section{Results}
\label{sect:results}

In this section, we evaluate the NN model performance by comparing the input versus the predicted SSP parameters. 


Fig. \ref{Age} shows the predicted versus the input ages for all galaxies up to magnitude $i$ $<$ 22 mag, as well as for galaxies with $i$ $=$ 16, 18 and 20. The agreement between the input and the predicted stellar age is excellent, with almost no average median bias ($\mu$=0.01 dex) and a small scatter ($\sigma_{NMAD}$=0.23 dex). The predictions tend to flatten towards old stellar populations (log($t$) $>$ 9.5 yr), as expected due to the small changes in the SED with age for such evolved galaxies. The predictions are least accurate for the youngest population interval (log($t$) < 7.5~yr, $\mu$=0.09 and $\sigma_{NMAD}$=0.43 dex), likely because two of the three libraries used for training lack such young stellar populations. We emphasize that these represent extremely young ages for typical observed galaxy populations. There is a significant dependence on the models performance with `observed' magnitude (i.e., S/N). The scatter values increase from $\sigma_{NMAD}$=0.13 dex at $i$ $ = $ 17 mag, up to  $\sigma_{NMAD}$=0.46 dex at $i$ $ = $ 21 mag. Note that the bias is quite constant in all the magnitude bins, although the flattening at high ages does become more evident for fainter galaxies.

\begin{figure*}
   \centering
   \includegraphics[width=0.49\hsize]{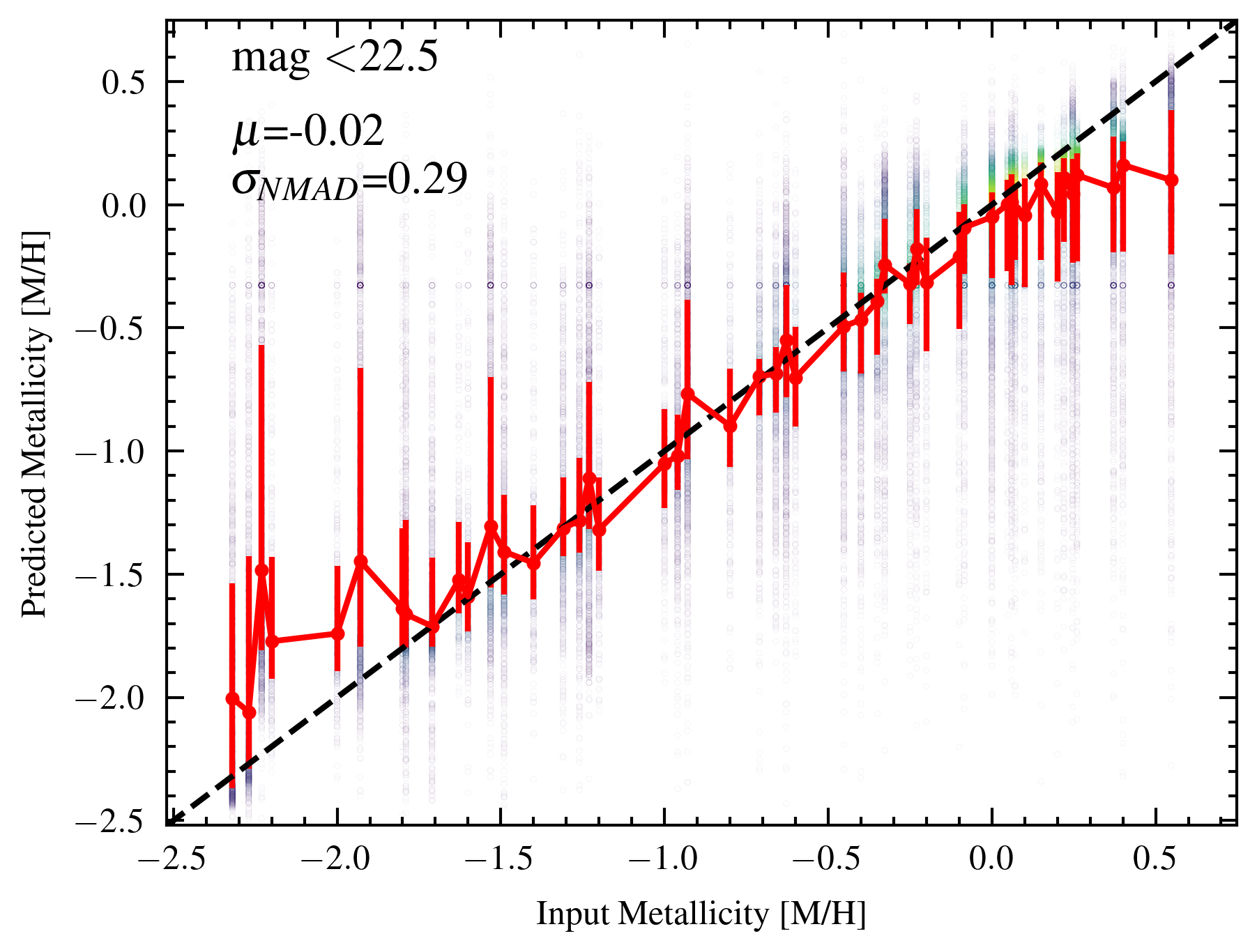}
      \includegraphics[width=0.48\hsize]{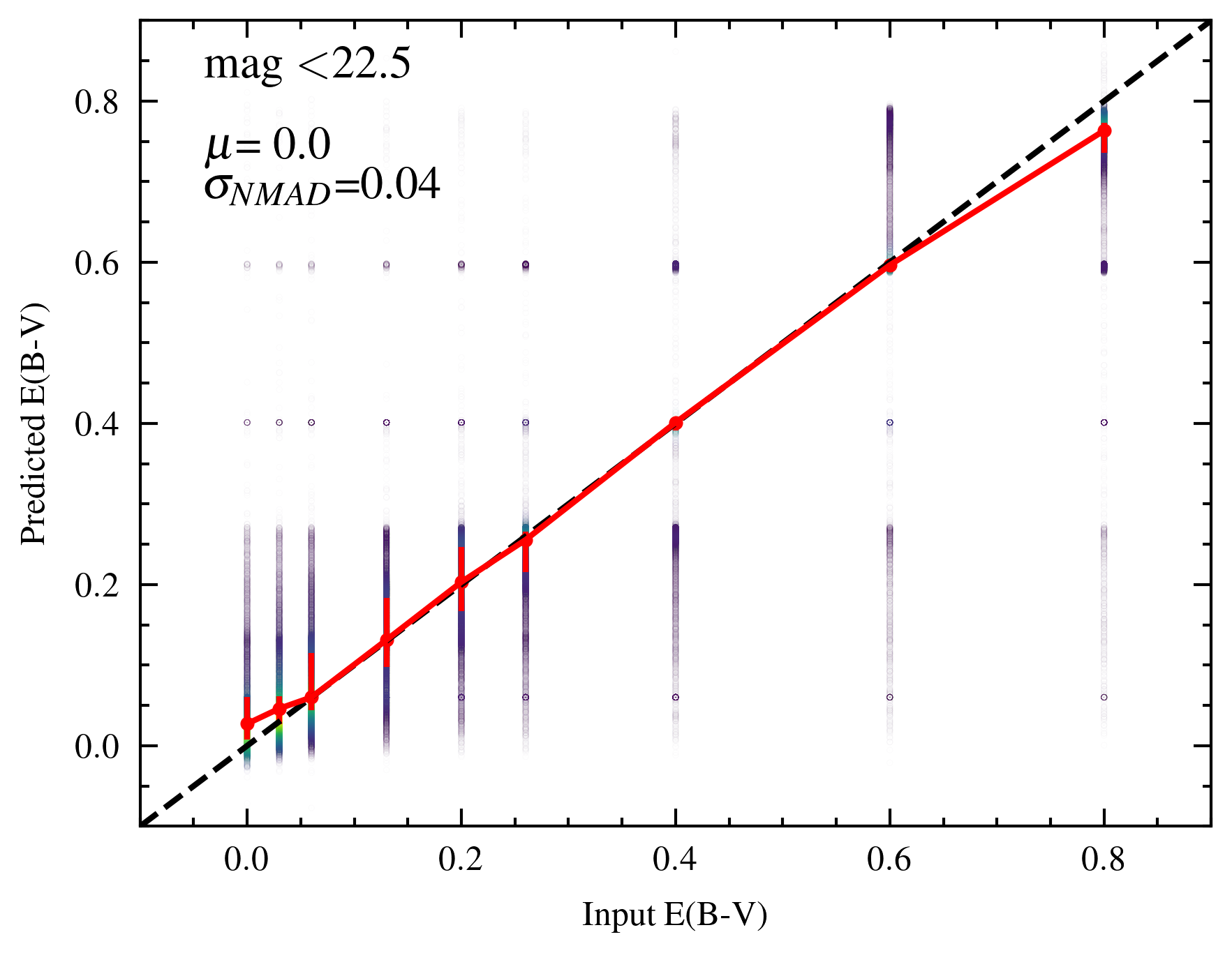}
   \caption{Same as Fig. \ref{Age} but for the [M/H] and E(B-V) estimates (left and right panels, respectively). }
    \label{Met-Ext}%
    \end{figure*}

A similar analysis is shown in Fig. \ref{Met-Ext} for the [M/H] estimates for all galaxies with $i$ $<$ 22.5 mag. The median bias and scatter are small ($\mu$=-0.02 dex,  $\sigma_{NMAD}$=0.29 dex). However, we note that there is a flattening of the predictions at the lowest and highest [M/H] values, i.e., the NN tends to overestimate the metallicity in the low-metallitcity regime, and underestimate it at the high-end. Fig. \ref{Met-Ext} also shows the comparison between the input and the predicted E(B-V) values. In this case, the results are excellent, with a  median bias close to zero and a small scatter ($\sigma_{NMAD}=0.04$ mag).


\subsection{Summary of the metrics}

\begin{figure*}
   \centering
   \includegraphics[width=\hsize]{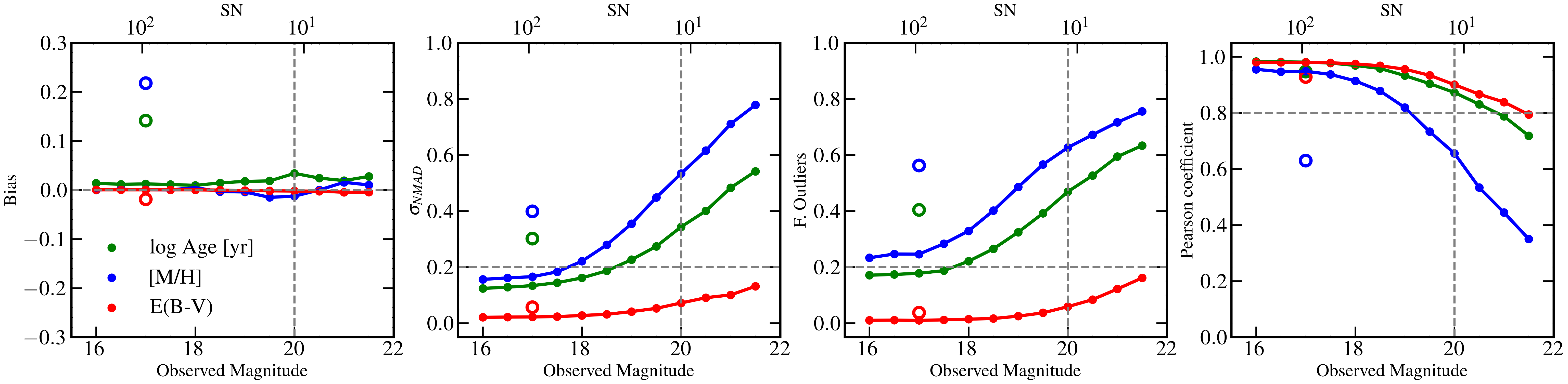}
   \caption{Summary of the metrics (from left to right, median bias, scatter, fraction of outliers and Pearson coefficient) in bins of of `observed' magnitude in the i-band (and average S/N in the narrow band filters) for the age (green), metallicity (blue) and dust attenuation (red) predictions. The empty circles are the metrics obtained by means of SED-fitting (details in Sect. \ref{sect:SED-fitting}) on the $i \sim$ 17 mag bin, highlighting the better performance of the NN estimates.}
    \label{Metrics}%
    \end{figure*}

We quantify the deviation of our predictions from the one-to-one relation by computing the Pearson correlation coefficient,  shown in Fig. \ref{Metrics} as a function of the observed magnitude (equivalent to S/N) for each of the SP parameters, together with the bias, scatter and fraction of outliers\footnote{Defined as those objects for which the difference between the input and the output is > 0.25 dex/mag}. As for the age estimates, there is a clear dependence of all these metrics on the S/N of the synthetic photometry. While this result may seem obvious, it is an important indication that our NN model is learning from the real signal (i.e., not overfitting), thus having more trouble with making predictions on noisy SEDs. Given the trends with `observed magnitudes', we conclude that our NN-model predictions are robust up to $i$ < 20 mag and might be used (with caution) for fainter galaxies.

Comparing the predictions of the different parameters, [M/H] shows the largest scatter, fraction of outliers and the smallest Pearson coefficient at all magnitudes, while E(B-V) seems to be the easiest parameter to recover, with a Pearson correlation coefficient above 0.8  and a fraction of outliers < 20\%, even for the faintest magnitude bins.

It is especially remarkable the ability of the NN to properly recover the dust attenuation, despite the lack of infrared coverage with J-PAS data. Dust attenuation primarily affects the blue–optical regime of galaxy SEDs, and the inclusion of near-infrared data is often advocated to help break the age–metallicity–dust degeneracy. While J-PAS does not include near-infrared photometry beyond 1$\mu$m, its contiguous narrow-band coverage and numerous age- and metallicity-sensitive features (e.g. the 4000 Å break, Balmer lines, and metal absorption features) allows us to recover robust estimates of dust attenuation. These results suggest that, despite the absence of infrared data, the essential information to constrain dust attenuation is effectively encoded within the J-PAS optical photometry.

\subsection{Degeneracies}

A well-known challenge when estimating SP parameters from photometry is the intrinsic degeneracy between age, metallicity, and dust attenuation, as all three affect the observed SED in similar ways —typically making it redder (see detailed discussion in \citealt{DiazGarcia2015}). While such physical parameter degeneracies can be alleviated using spectral features in spectroscopic data (e.g., \citealt{DS2019b, DS2020, Bernardi2019}), they are more difficult to address when using photometry, since the filter convolution washes-out those small differences.  In SED-fitting methods, all SP parameters must be fitted simultaneously. This requires identifying the spectral template — characterized by a specific age, metallicity, and dust attenuation — that best reproduces the data. Selecting a template with an age that is too old may require compensating by choosing a lower metallicity or a less dust-attenuated model; otherwise, the resulting spectrum would appear too red. One of the key advantages of the methodology presented in this work is that these three parameters are estimated independently using NNs trained specifically to recover each parameter, without the need to fix the others. In other words, the NN that predicts age is agnostic to the values of metallicity and dust attenuation predicted by the other NNs. As a result, the traditional notion of parameter degeneracy - i.e., the correlation between errors in the predicted quantities - is largely mitigated in the context of our approach.

\begin{figure*}
   \centering
   \includegraphics[width=0.32\hsize]{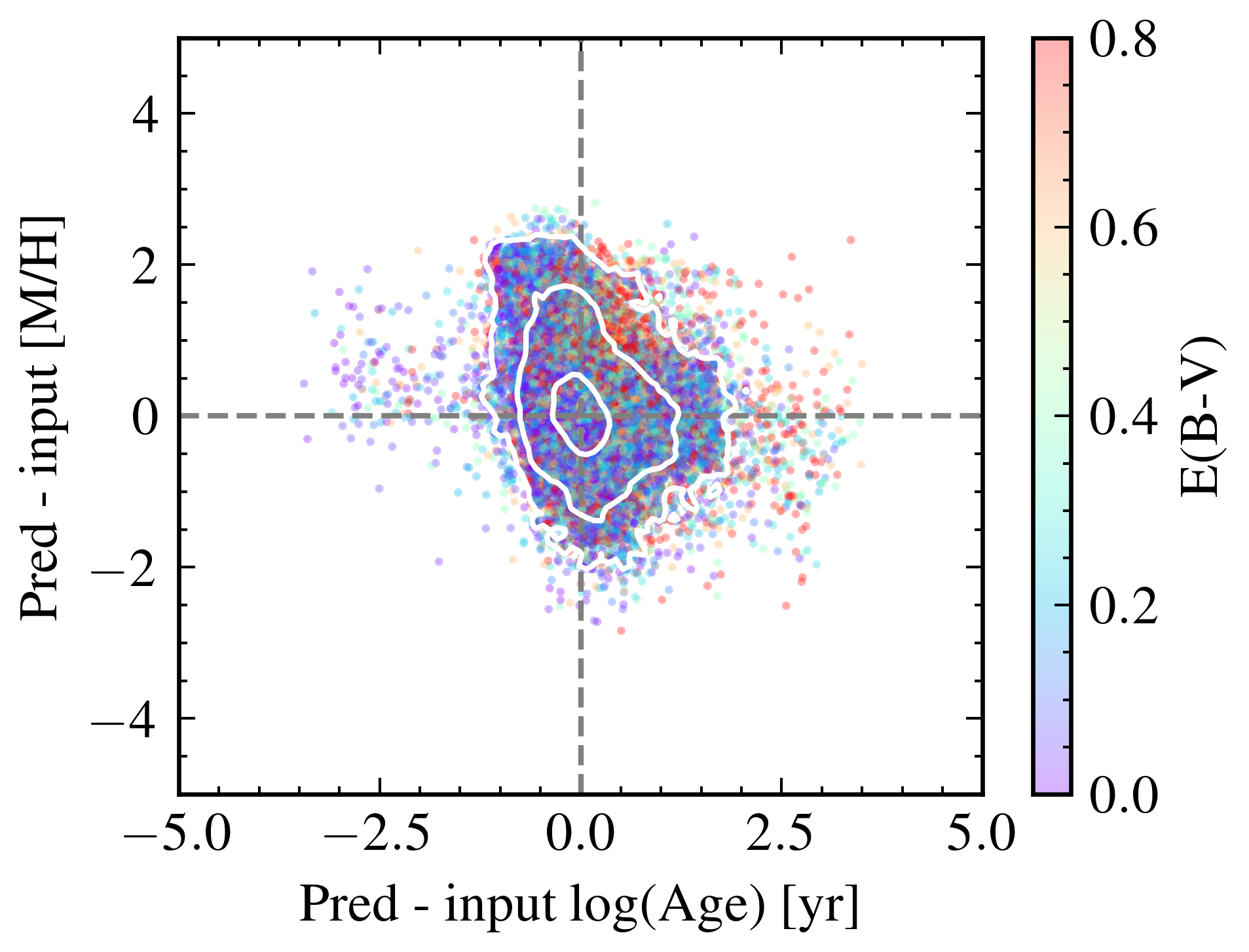}
    \includegraphics[width=0.34\hsize]{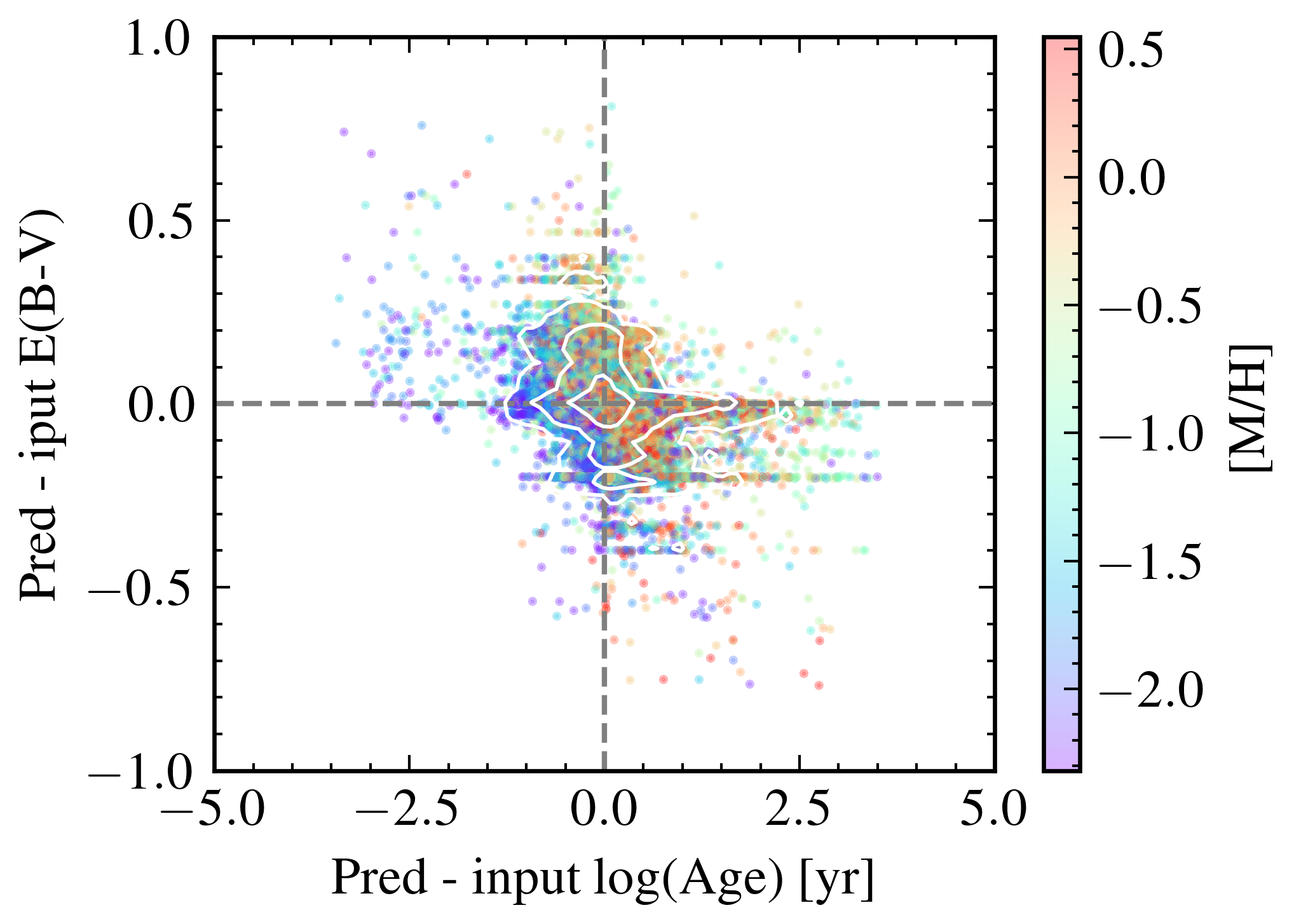}
    \includegraphics[width=0.32\hsize]{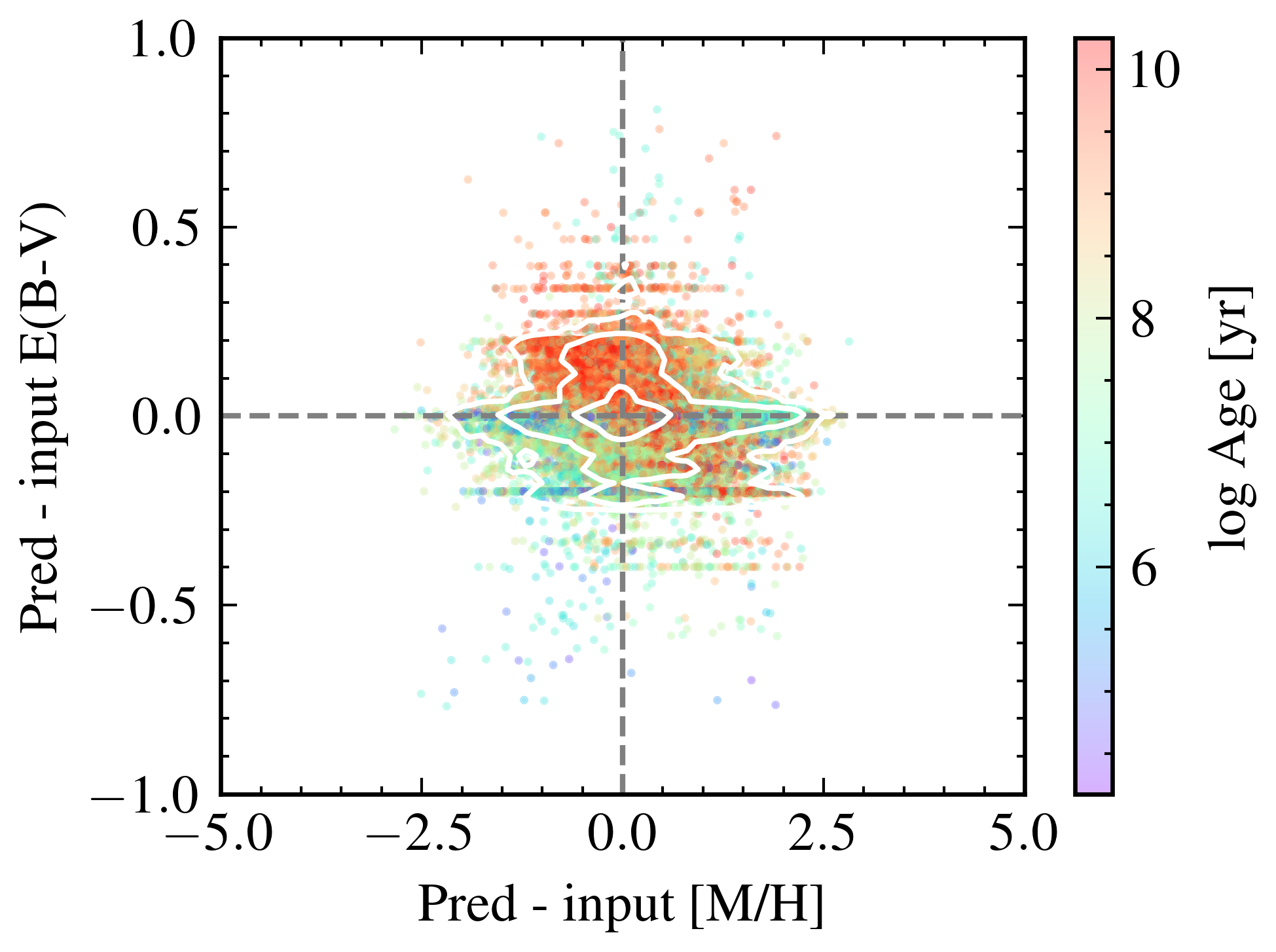} 
\includegraphics[width=0.32\hsize]{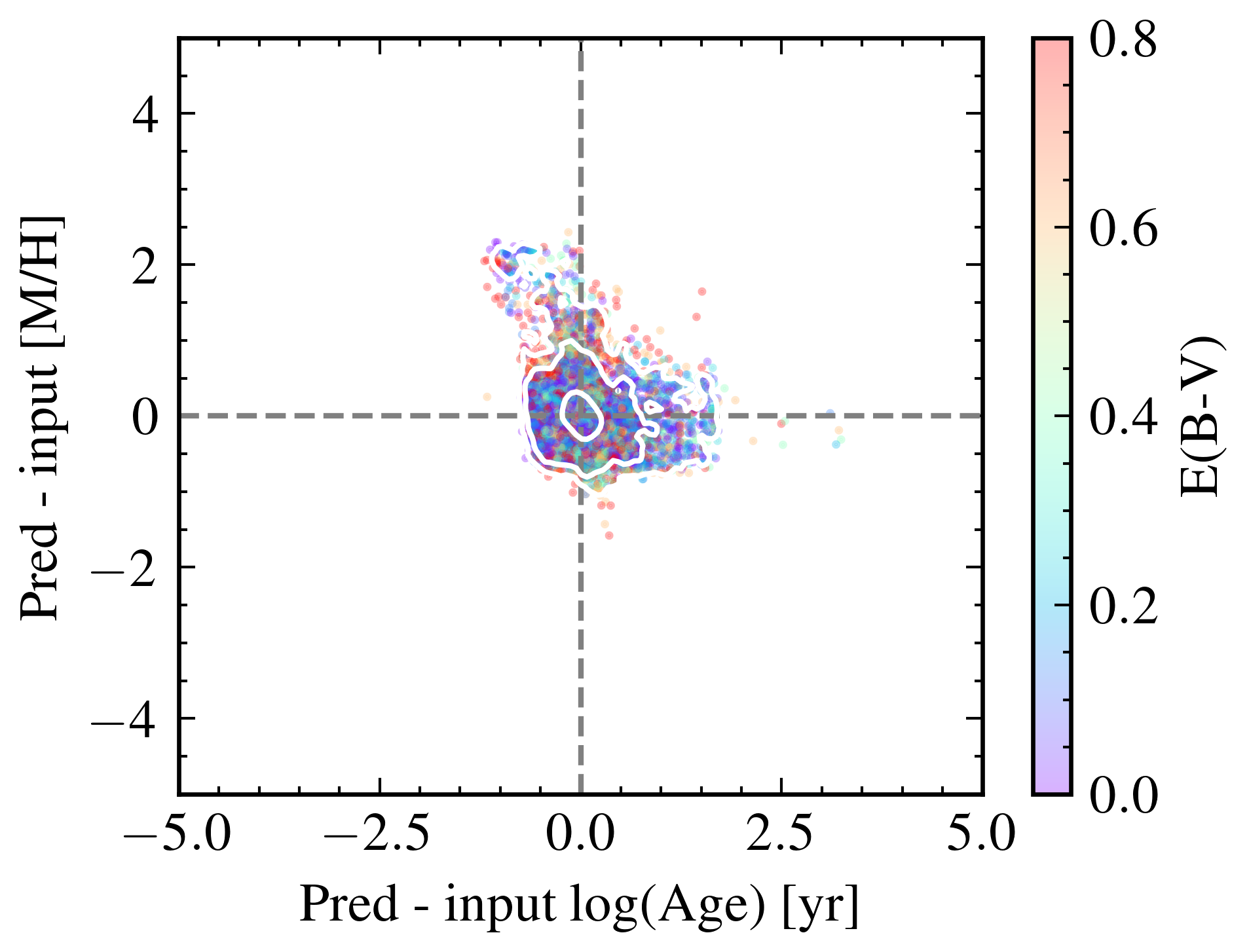}
    \includegraphics[width=0.34\hsize]{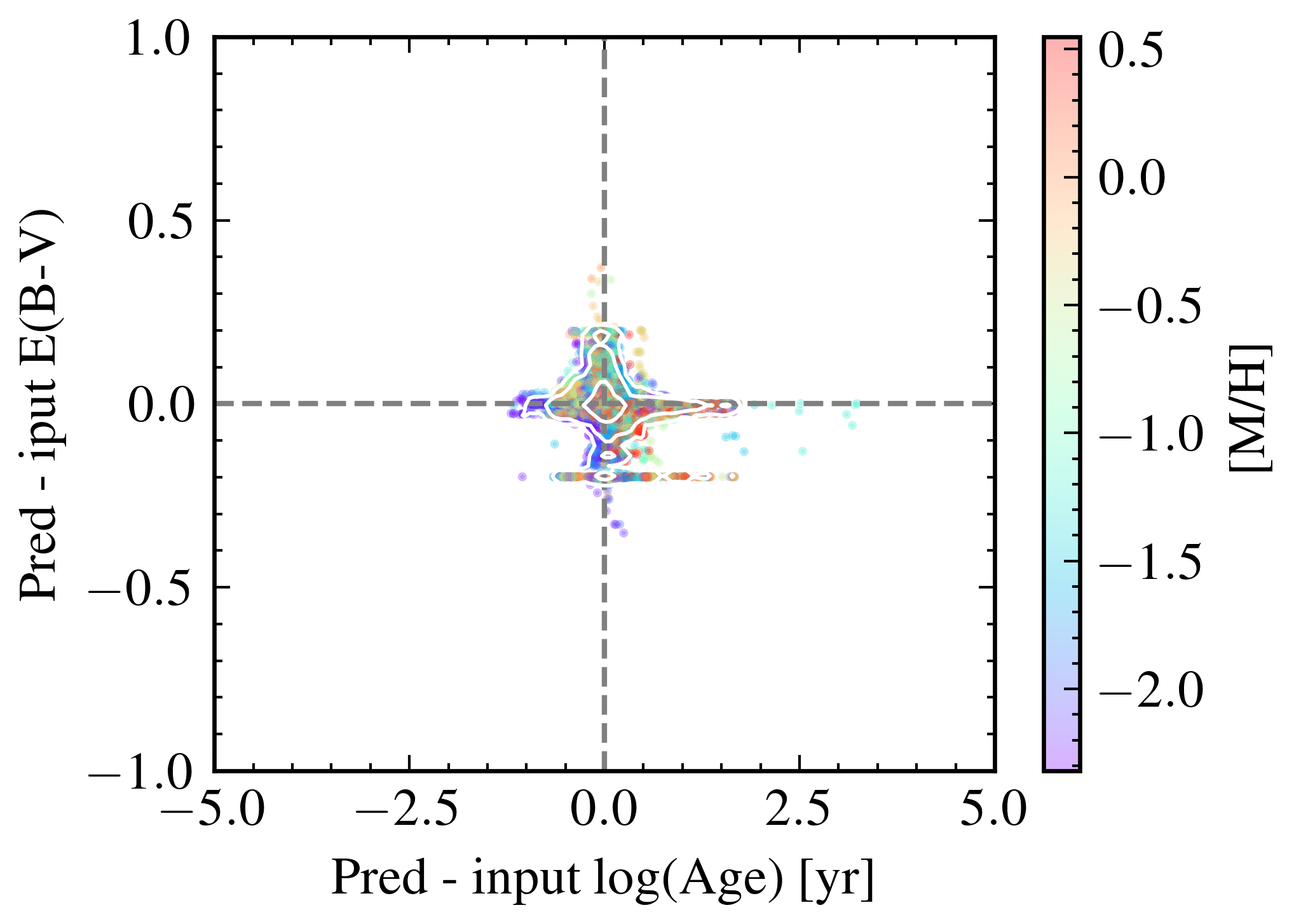}
    \includegraphics[width=0.32\hsize]{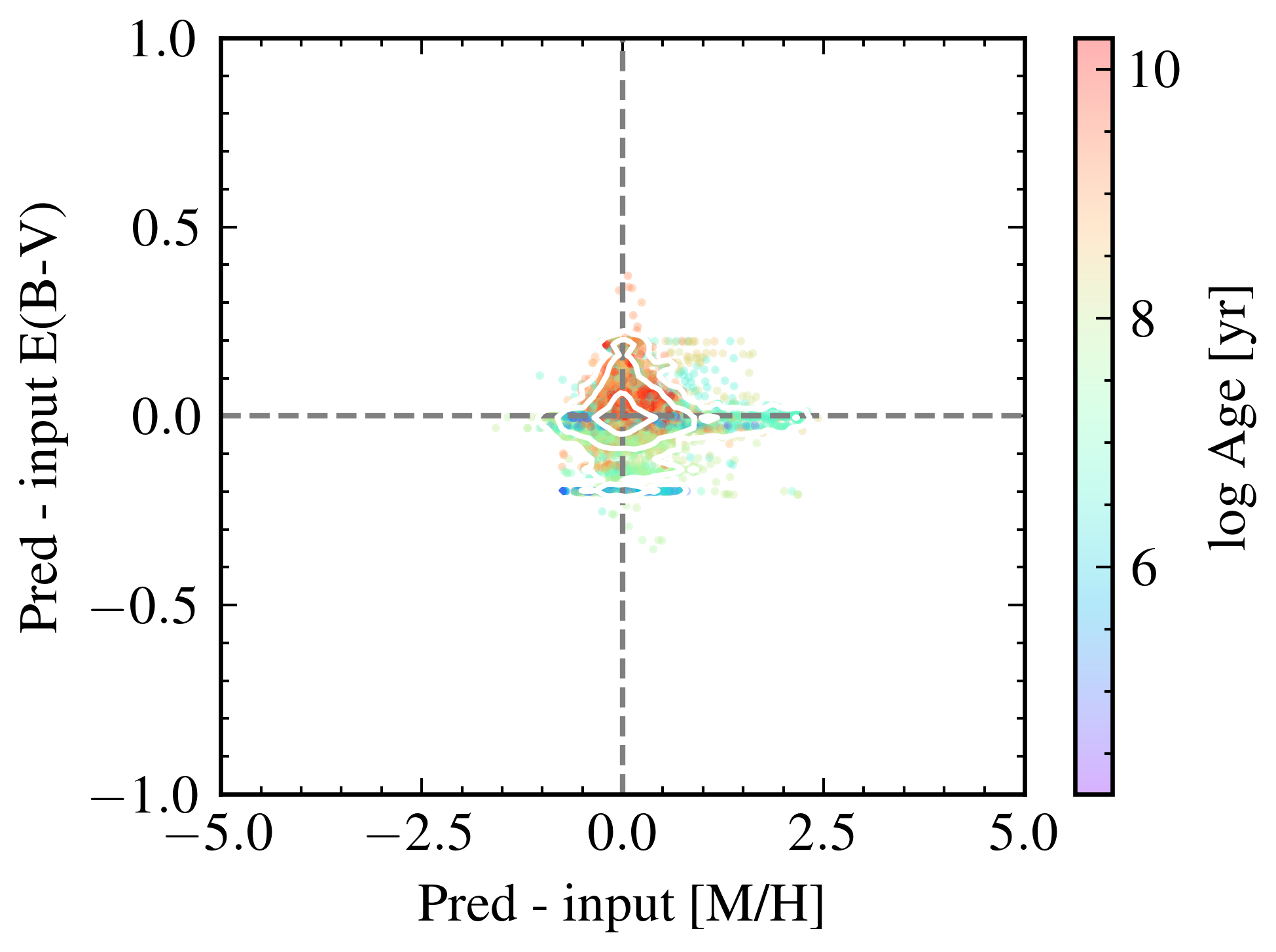} 
    
   \caption{Left: Difference in the predicted  metallicity  versus  age difference, color coded by dust attenuation. Middle:  Difference in the predicted  dust attenuation  versus  age difference, color coded by metallicity.  Right:  Difference in the predicted  dust attenuation  versus  metallicity difference, color coded by age. Upper panels show  galaxies with $i$ < 20.5 mag, bottom panels galaxies with $i$ < 17.0. The white lines show 1, 2, and 3$\sigma$ contours.}
              \label{degeneracies}
    \end{figure*}

To study possible correlations among the derived stellar population parameters, Fig. \ref{degeneracies} shows the parameter space of the differences between the input and predicted parameters across all possible configurations. The data distribution look uniform, with no evident structure that would indicate correlations in the parameter estimate errors. In particular, the age–metallicity parameter space is well behaved—centered around (0,0) —and shows no clear dependence on dust attenuation. When E(B–V) is shown on the y-axis, some structure emerges in the form of horizontal bands, related to the discrete sampling of E(B–V) values in the SSP models, but not due to degeneracies in the SSP estimates. Using a finer E(B–V) grid in the training and test samples would likely reduce this effect. A weak residual trend is visible in the E(B-V) vs age plot - where metallicity increases from left to right - and in the E(B-V) vs [M/H] plot - where age increases from bottom to top. These patterns are more prominent outside the 2$\sigma$ contours and are mostly driven by the faintest galaxies, suggesting mild degeneracies between age, metallicity, and dust attenuation in the low-S/N regime. 

Indeed, the bottom panels of Fig. \ref{degeneracies} show the residuals obtained for the bright sample  ($i<17$ mag) where these trends are even less pronounced, highlighting a notable contrast with the parameter estimate degeneracies typically seen in SED-fitting, which affect all magnitude ranges (see discussion in \citealt{DiazGarcia2015}). This behavior suggests that, despite the very similar SED shapes, there is likely a subtle signal in the data that the NN is able to capture, allowing it to accurately recover the ground truth values. This highlights one of the key advantages of NN over traditional SED-fitting methods.

\subsection{SSP Model dependence}

In this work, we combine synthetic photometry from different SSP libraries to train the NN models. After several tests, we found that the network performs more robustly when information from multiple libraries is mixed. Additionally, incorporating different IMFs, available in the E-MILES SSPs, further enhances the NNs performance. This can be interpreted as a form of data augmentation, helping the network focus on general trends rather than specific features of individual libraries or IMFs.

\begin{figure*}
   \centering
   \includegraphics[width=\hsize]{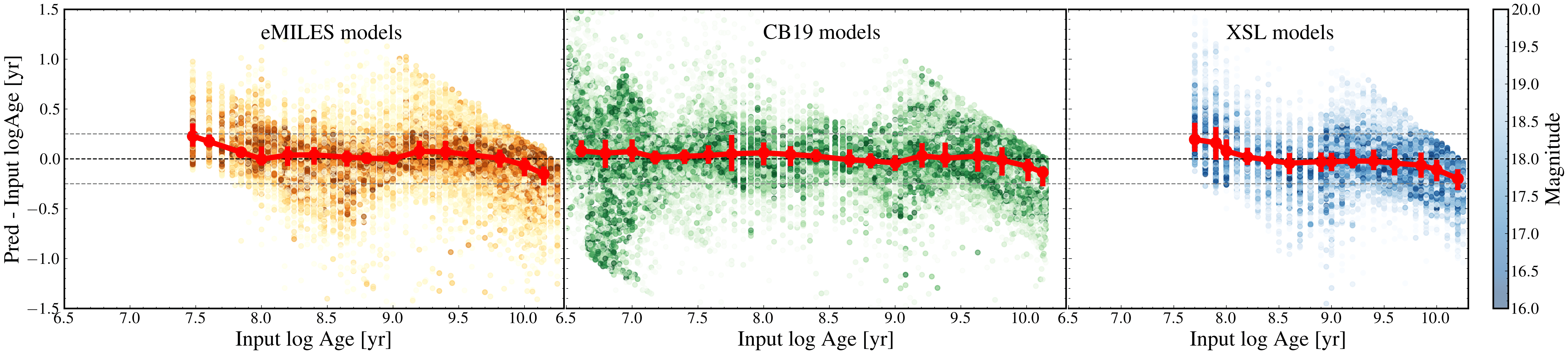}
      \includegraphics[width=\hsize]{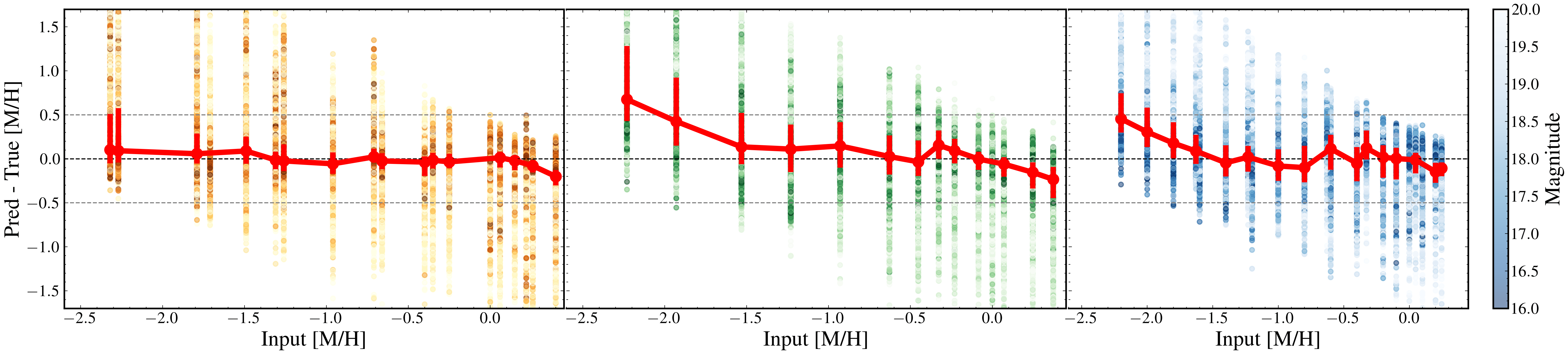}
         \includegraphics[width=\hsize]{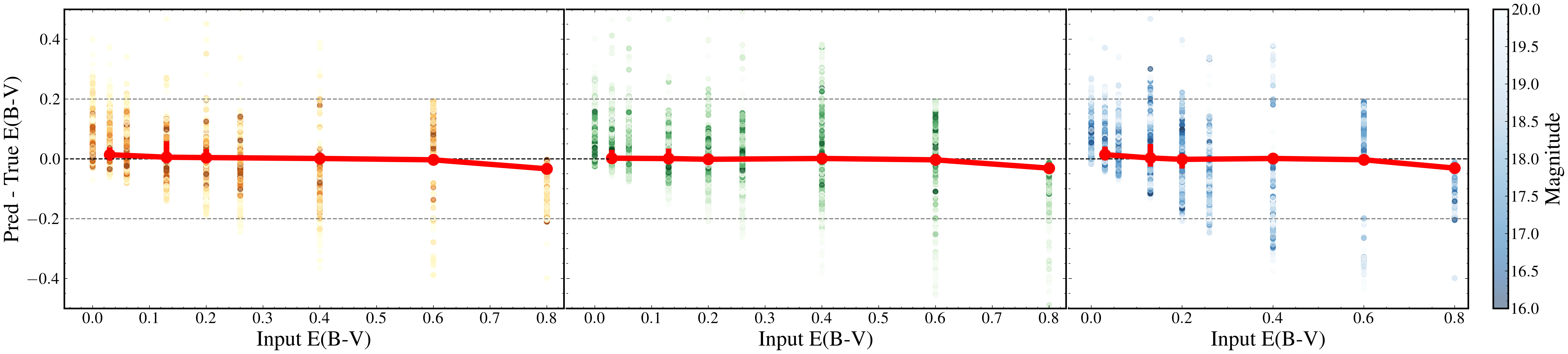}
   \caption{Difference between the predicted and the input parameters (from top to bottom, logAge, [M/H] and E(B-V)) for the sample limited to  $i$ < 19.5, divided into different SSP synthetic photometry (from left to right: E-MILES, CB19 and XSL). The color intensity corresponds to the `observed' magnitude. Red points indicate the median values in bins, with red error bars showing the interquartile range (25th to 75th percentiles) of the distribution.}
              \label{SSP-model}%
    \end{figure*}

Since we combined the three SSP libraries in our training sample, we explore if there is any dependence of our results on the SSP library for which the predictions are made. For that purpose, Fig. \ref{SSP-model} shows the difference between input and output parameters for the three SSP libraries used. In general, all the SSP models behave very well, with almost no bias in all the estimated range. It is worth noting, that the metallicities are overestimated for galaxies with [M/H] < -1.5 for the CB19 and XSL models. Also, the ages are slightly overestimated for the youngest SSP models for the E-MILES and XSL libraries. On the contrary, there is no  age bias for bias for the CB19 SSP models, nor metallicity bias for E-MILES. We recall the reader that all the templates younger than 30Myr come  precisely from the CB19 library, while the lowest metallicity values are those of the E-MILES library, so these results are not surprising.

The small oscillatory features visible in Fig. \ref{SSP-model} probably arise as a result of the division of the test sample into three subsets, one for each SSP library, which reduces the statistics and enhances local fluctuations. Since the NN is trained simultaneously with all libraries, these patterns may also trace minor systematic differences among the SSP models (e.g. stellar libraries or isochrone choices). The effect is mainly seen in the individual data points, while the median residuals remain smooth and close to zero, confirming that no significant systematic bias is present in the model predictions.

\section{Discussion}
\label{sect:discussion}

Predicting SP parameters from photometry is an extensively  investigated topic, and doing a full review is not the purpose of this work. Making a fair comparison of the performance of our results with other methodologies such as SED-fitting is not trivial, given the differences in the photometry, stellar population models, and parametrization of the star formation histories. In this section we compare our results with those obtained by SED-fitting on the same synthetic photometry test sample, and compare our results with those in the literature whose analysis is somehow comparable to the analysis presented here.

\subsection{Comparison with SED-fitting}
\label{sect:SED-fitting}

The most common approach to derive SP parameters from photometry is by means of SED-fitting. To compare our methodology with SED-fitting procedure in a fair way, we fit the exact same synthetic photometry of the test sample using the CB19 models with \textsc{Cigale} SED-fitting code. To be more conservative, we only do this exercise for the test sample coming from the CB19 library. We assume the star formation histories to be similar to SSPs (i.e.,  a single burst very short formation timescale, $\tau$ = 1 Myr) with a fine grid of possible age, metallicity\footnote{note that we use all the allowed metallicity values in CIGALE for these excersise} and dust attenuation values, as described in Table \ref{tab:SED-fitting}, no nebular lines and \cite{cardelli+89} attenuation law. The results are shown in Fig. \ref{Metrics} as empty circles. For simplicity, we limit our analysis to  photometric errors typical from mag $\sim$ 17 galaxies. We obtain median bias, scatter and percentage of outliers $\mu$ = (0.14 dex, 0.22 dex, -0.02 mag), $\sigma_{NMAD}$ = (0.30 dex, 0.40 dex, 0.06 mag), 
f$_{o}$ = (40\%, 56\%, 4\%)   at $ i \sim$17 mag for age, metallicity and dust attenuation, respectively. It is clear that,  for the three SP parameters, all the metrics reported are worse for SED-fitting estimates (the detailed one-to-one comparison is shown in Fig. \ref{fig-App:SED-fittig}). This confirms that NN are a very powerful tool for inferring the SP properties of galaxies and encourages us to continue exploring their use, despite the limitations discussed in Sect. \ref{sect:conclusions}. 

\begin{table*}[h]
    \centering
    \begin{tabular}{l l}
        \textbf{Parameter} & \textbf{Possible values} \\
          \hline
        Metallicity [Z] & 0.0001, 0.0002, 0.0005, 0.001, 0.002, 0.004, 0.006, 0.008, 0.01, 0.014, 0.017, 0.02, 0.03, 0.04, 0.06 \\
        Age [Myr] & 1, 3, 5, 8, 10, 20, 30, 40, 50, 80, 100, 500, 800, 1000, 2000,3000, 4000, 5000, 6000, 7000, 8000, 9000, 10000, 11000 \\
        E(B-V)$_{lines}$ & 0.0, 0.1, 0.2, 0.3, 0.4, 0.5, 0.6, 0.7, 0.8, 0.9, 1.0, 1.3, 1.5, 1.9, 2.2 \\
    \end{tabular}
    \caption{Parameter space covered by the CB19 models used in the SED-fitting procedure with CIGALE. Note that we use all the possible metallicity values.}
    \label{tab:SED-fitting}
\end{table*}

\subsection{Comparison with literature}
\label{sect:literature}

\citet{GonzalezDelgado2021} compare the results of the parametric SED-fitting code \textsc{BaySeAgal} on the mini-JPAS sample with the output of three non-parametric  SED-fitting codes: \textsc{Muffit}, \textsc{AlStar}, and \textsc{TGASPEX} \citep{Magris2015}. They concluded that, given the differences in the different SED-fitting approaches, the precision of the SP properties depends on the S/N and the property with the largest uncertainty is the metallicity, in agreement with our results.

In \cite{MejiaNarvaez2017} the authors report results of SED-fitting on mock observations from the Synthetic Spectral Atlas of Galaxies, including spectroscopy, narrow-band photometry and broad-band photometry. The narrow-band photometry is aimed to reproduce J-PAS observations with different levels of noise and the SP parameters are estimated using the \textsc{DynBas} non-parametric spectral fitting code \citep{Magris2015}. They report values of bias and scatter of $\mu$ =(-0.01, -0.01, -0.016) $\sigma$ = (0.2, 0.17, 0.2) for $\log t$, $\log Z$ and $A_v$, respectively, comparable to our results. However, we should note that their mock observations include star formation histories more complex than SSP, and  the library used in their fitting is \cite{BC2003}. \cite{Liew-Cain2021} trained a convolutional neural network  on synthetic  J-PAS photometry from ~21000 spectra of a sample of ~200 galaxies from the CALIFA survey \citep[CALIFA]{Sanchez2012} with m$_B \sim$18 mag. The authors assumed as `true' metallicity and age values those obtained by means of full-spectral fitting on the emission-line cleaned spectra with \textsc{STECKMAP} code \citep{Ocvirk2006}. They report a very small bias and a standard deviation $\sigma \sim 0.2$ dex for both log(age) [yr] and Z, comparable to our results (see Fig. \ref{Metrics}), despite the fact that the age and metallicity parameter space targeted in that work is much more limited than ours (log$t$ [yr] = [9.2, 10] and [M/H] = [-0.6, 0.2]). In a recent work, \citet{Wang2024} train a Convolutional NN to predict SP parameters from Sloan Digital Sky Survey (SDSS) spectra, using as `labels' the values obtained by full spectral fitting with \textsc{pPXF}. They obtain a scatter $\sigma \sim$ 0.11 for the age and metallicity and  $\sigma \sim$ 0.018 for the E(B-V). Although these values are smaller than the ones we obtain, we note that those are estimated using spectroscopic information for a bright sample of galaxies ($r \sim$ 18). We would like to emphasize that none of the aforementioned studies compares their results to the ground truth, as we do in this work. Instead, they assess performance against SP parameters derived from alternative methods such as spectral fitting, which inherently carry their own uncertainties and biases that can propagate into the evaluation of the results.

In a comprehensive study that uses information from simulations as ground truth, \cite{Woo2024} compare the performance of several popular spectral fitting codes with that of a CNN (\textsc{StarNet}) using mock spectra from IllustrisTNG-100 simulations \citep{Nelson2019} mimicking SDSS.  Although their results are not directly comparable to ours given the large differences in S/N and spectral resolution, they find that \textsc{StarNet} vastly outperforms conventional codes, supporting our findings on the power of machine learning based methods for deriving SP parameters.

\subsection{Caveats}\label{sec:caveats}

Although the results are very promising, there are a few caveats to our current approach that should be noted. First, the SSP models used to train the neural network are all at redshift z = 0. Given the narrow width of the J-PAS filters ($\sim 150 \AA$), the redshift bins required to accurately sample the synthetic photometry are approximately $\Delta$z $\sim$ 0.03. While generating synthetic photometry up to z $\sim$ 1 is feasible, it would significantly increase the computational cost of our method, and we leave this for future work. Additionally, as previously mentioned, our training sample does not include emission lines or complex star formation histories, very common in the local Universe. The performance of the neural network when applied to composite stellar populations (CSPs) is discussed in Appendix \ref{Appendix:CSP}. As expected, the accuracy of the predicted properties decreases with the complexity of the CSP, with the poorest results obtained for models with a constant star formation history. This is in agreement with previous work (e.g., \citealt{Breda2022}) showing that age estimates  are biased when nebular continuum emission is not taken into account. An alternative would be to use simulation based inference on non-parametric SFHs, as done in \cite{IglesiasNavarro2024} for early-type galaxies with spectroscopic data. Finally, the current approach is not optimised for missing data. Since the NN is trained in fluxes, missing data in the strict sense do not occur. Issues such as masking or bad pixels could be overcome by linearly interpolating neighboring filters, which preserves the the accuracy of the derived parameters as long as the fraction of missing bands remains small (Hernán-Caballero et al. in prep.). The number of J-PAS sources in the public EDR  with observations in all bands is 71\%.

We leave these improvements for future work and present our results as a promising proof of concept, highlighting the power of combining J-PAS narrow-band photometry and state-of-the-art synthetic stellar population libraries with deep learning inference.

\section{Test on J-PAS galaxies}
\label{sect:JPAS}

\begin{figure*}
   \centering
    \includegraphics[width=0.32\hsize]{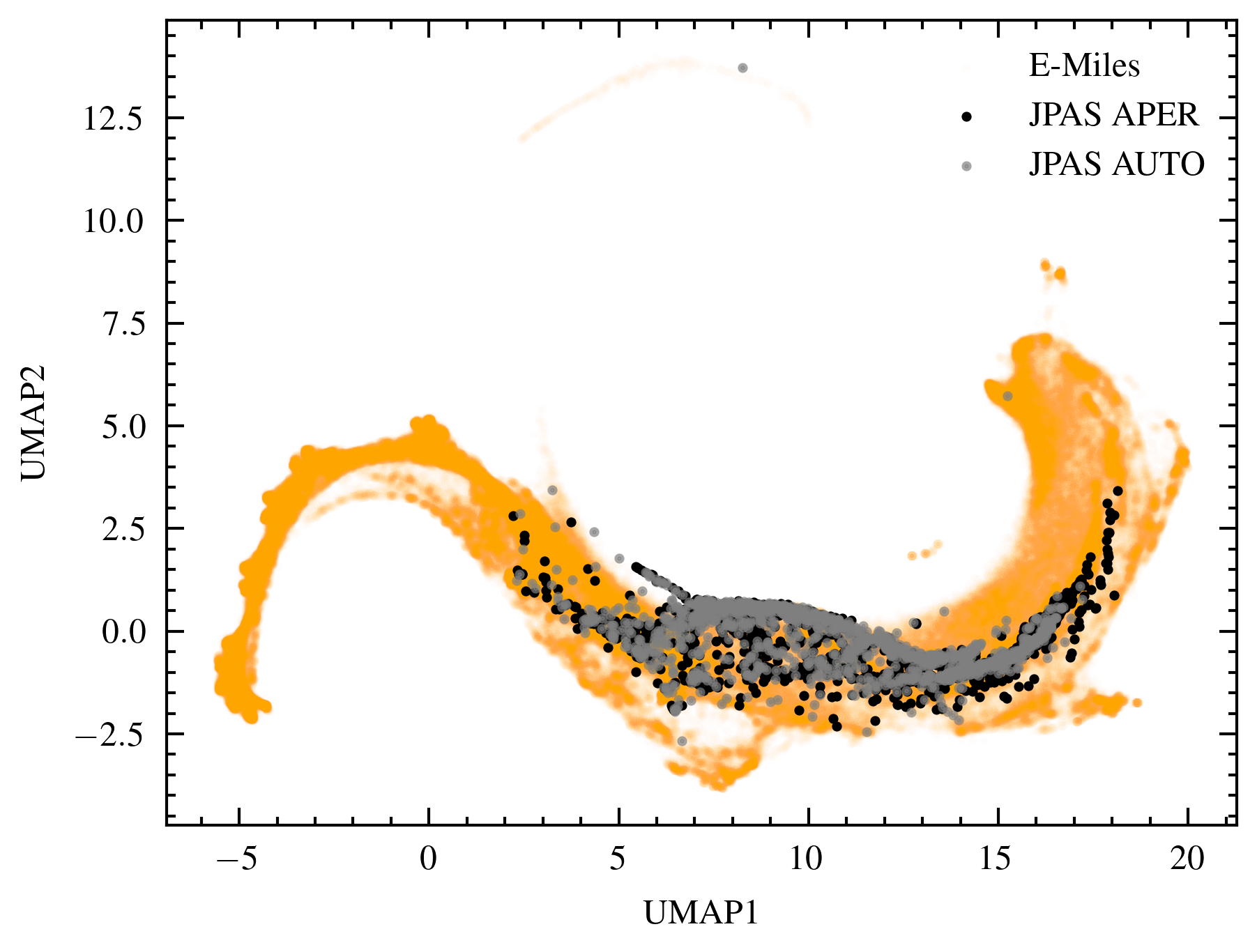}
      \includegraphics[width=0.32\hsize]{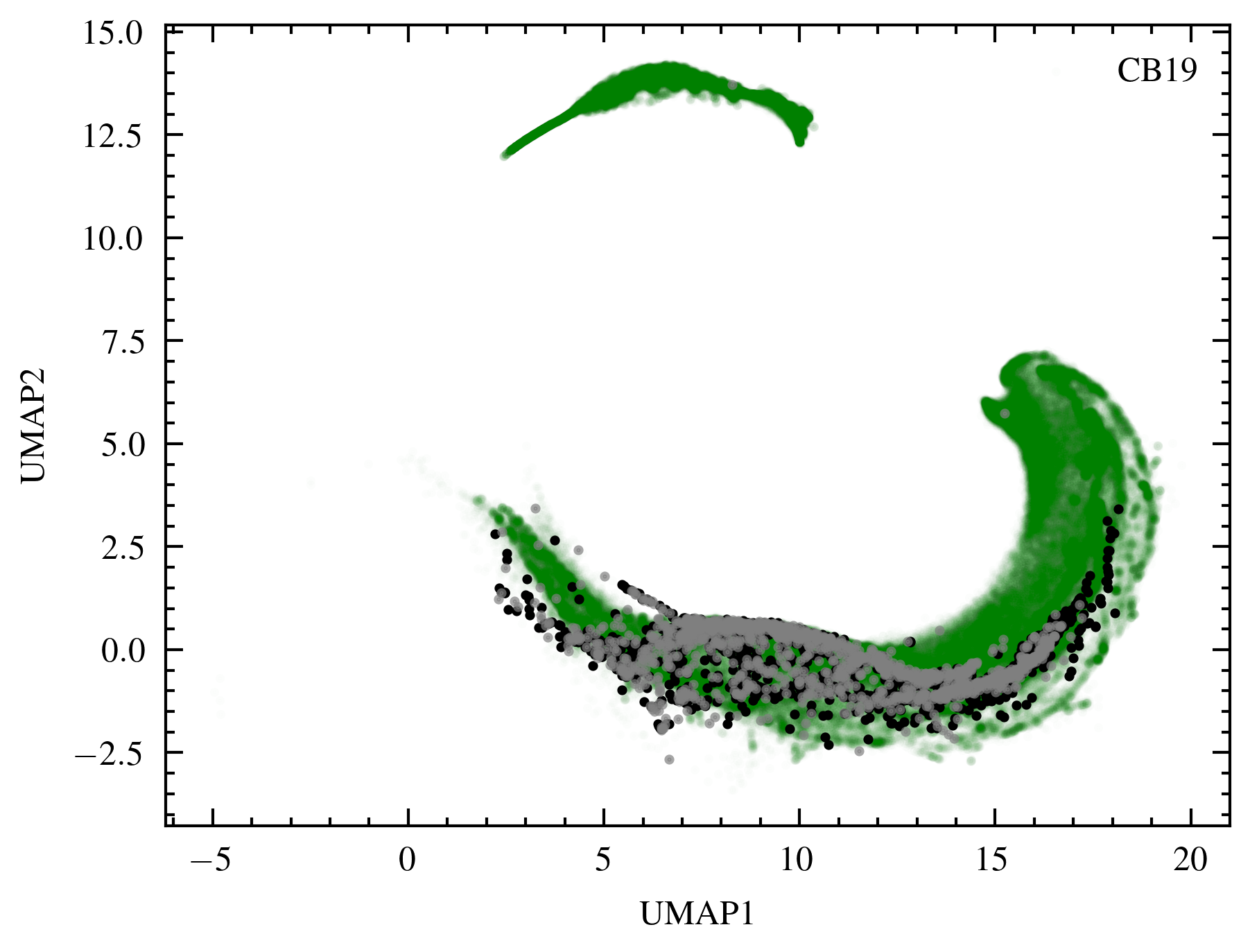}
         \includegraphics[width=0.32\hsize]{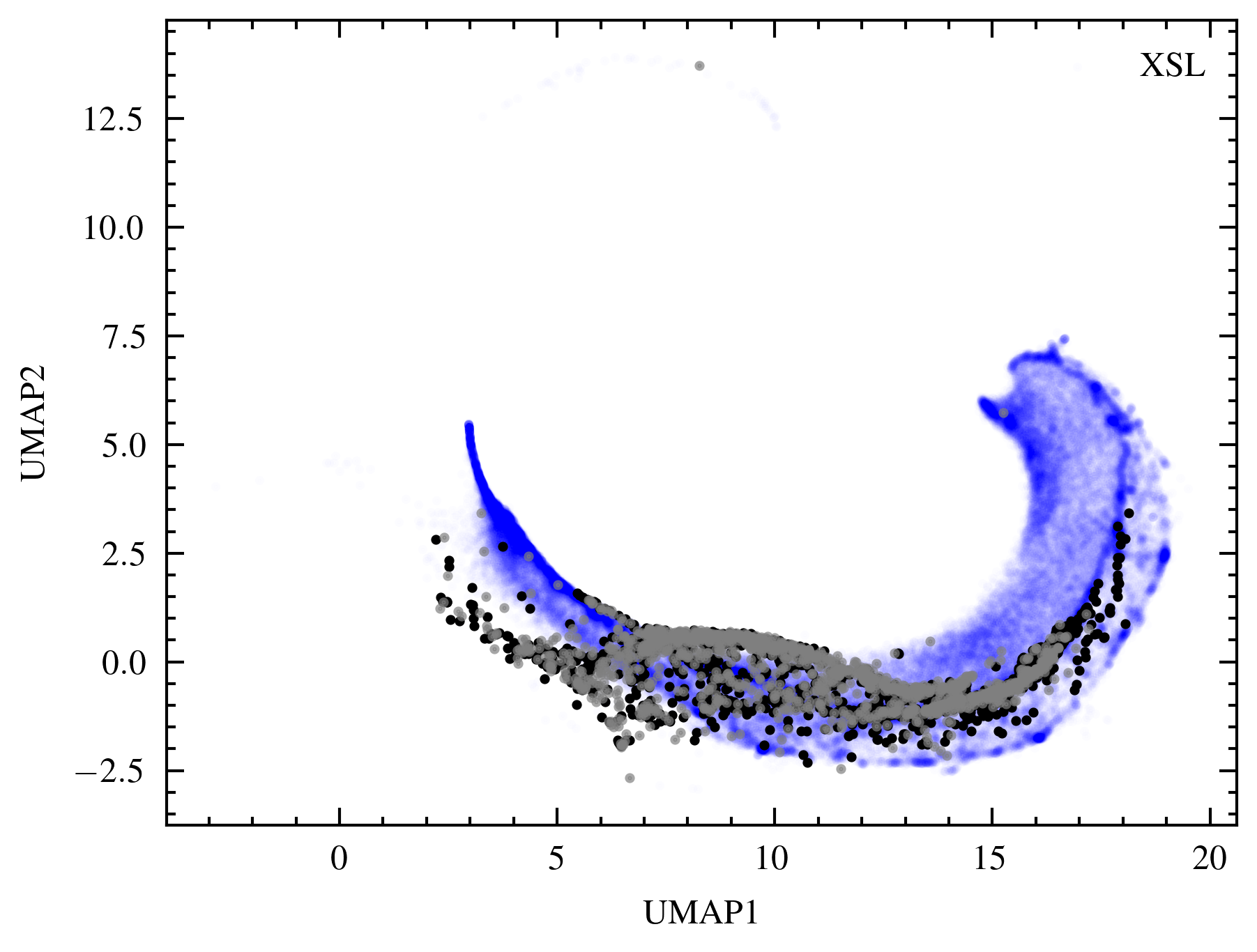}
   \caption{Projection of the synthetic photometry and the real J-PAS photometry in the 2D parameter space obtained by UMAP dimensionality reduction algorithm. Black and gray data show the J-PAS APER$\_$COR$\_$3$\_$0 and AUTO fluxes, respectively, while each panel shows the synthetic photometry for the three SSP libraries used in the training of the NN (E-MILES in orange, CB19 in green and xsl in blue).  While a single library does not fully cover the parameter space observed in real J-PAS data, the coverage is notably improved when the three libraries are combined together.}
              \label{UMAP}%
    \end{figure*}

The first J-PAS data release was made public in 2024 (Vázquez Ramió in prep.). We use the released J-PAS photometry to compute the stellar population properties for a sample of local galaxies from the NN predictions. We have made a selection of low-z galaxies with spectroscopic redshift $zspec < 0.1$ (and $|zphot-zspec|$< 0.005)\footnote{We require $zphot$ and $zspec$ to agree as a way to confirm that the spectroscopic redshift is correct and that there are no strong artifacts in the photometry that would affect the SP estimates.}, $i < 20$ mag, resulting in 1478 galaxies. We use as input for the NN  predictions the APER$\_$COR$\_$3$\_$0$\_$FLUX values\footnote{The results are not severely affected when using AUTO photometry}, normalized by their mean. In Fig. \ref{UMAP} we present the projection of both synthetic and real photometry onto a 2D parameter space using UMAP \citep{mcinnes2020}, a non-linear dimensionality reduction technique that preserves both local and global structure, making it well-suited for visualizing complex, high-dimensional datasets in 2D.

The overlap between the real observations and the synthetic photometry in the 2D UMAP projection suggests that both datasets span a similar region of parameter space. While 2D projections cannot capture all aspects of the high-dimensional structure - so it does not guarantee 
the consistency between the modeled and real data distributions - , when using only a single SSP library, the synthetic models fail to fully cover the range of colors observed in the J-PAS data, implying a genuine mismatch in the original, higher-dimensional space. On the other hand, when combining the three SSP libraries, the synthetic photometry achieves significantly better coverage, aligning more closely with the observed data. This supports the strategy of integrating all three libraries into the training sample, as it enhances the robustness and generalization of the neural network.

  We apply the NN models trained on synthetic photometry to the J-PAS observations and we compare our results with those derived from SED-fitting using \textsc{MUFFIT} with two-burst CSP with CB17 models and Chabrier IMF (more details in Díaz-García, in prep.). For this excercise, we use a more restrictive criteria for the redshift limit ($z < 0.03$)  and for the photometric quality, avoiding galaxies close to bright stars (FLAGS$\_$MASK = 2) or with all the bands having bad FLAGs, resulting in a sample of 90 galaxies.

\begin{figure*}
   \centering
       \includegraphics[width=0.33\hsize]{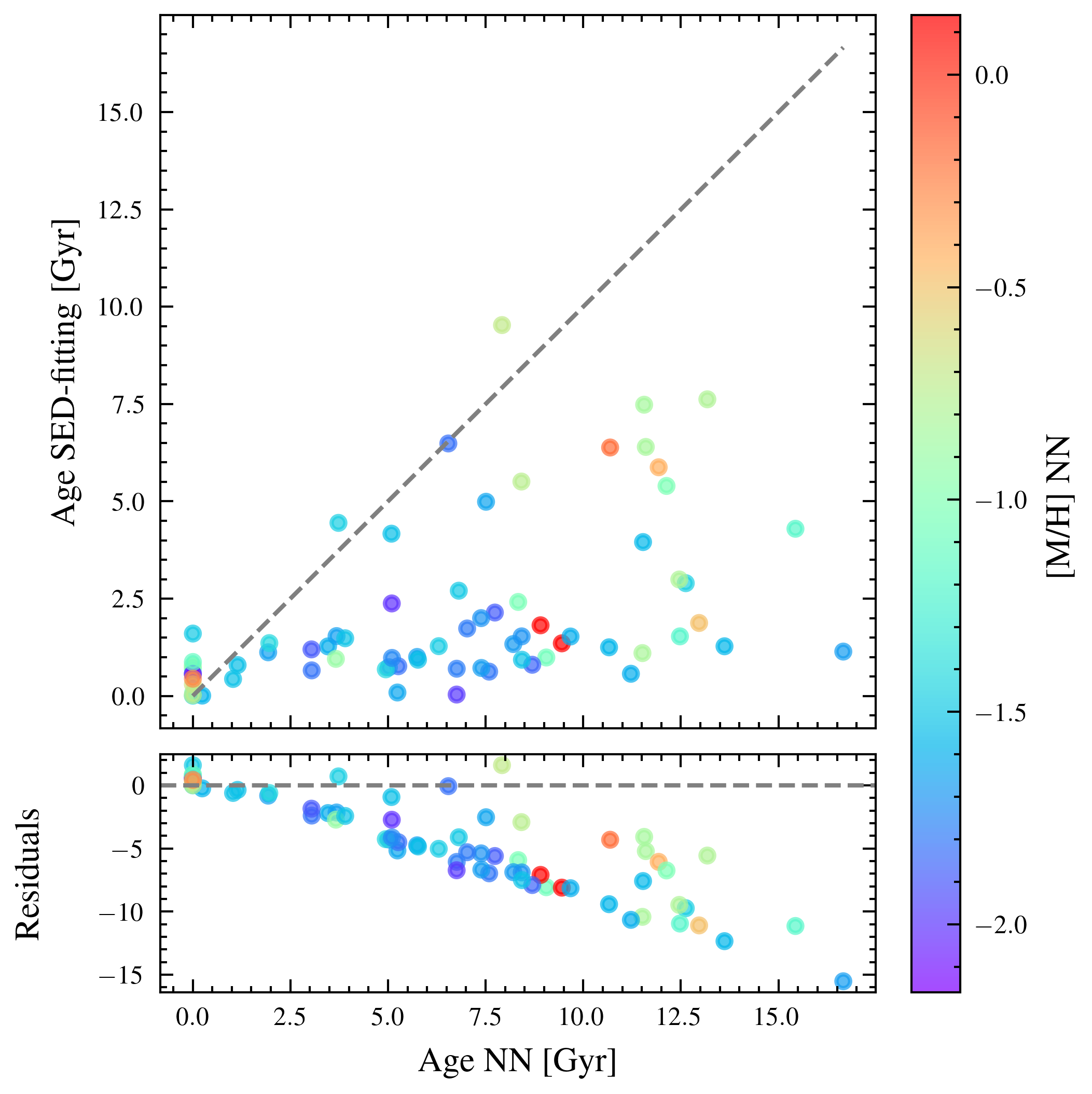}
    \includegraphics[width=0.33\hsize]{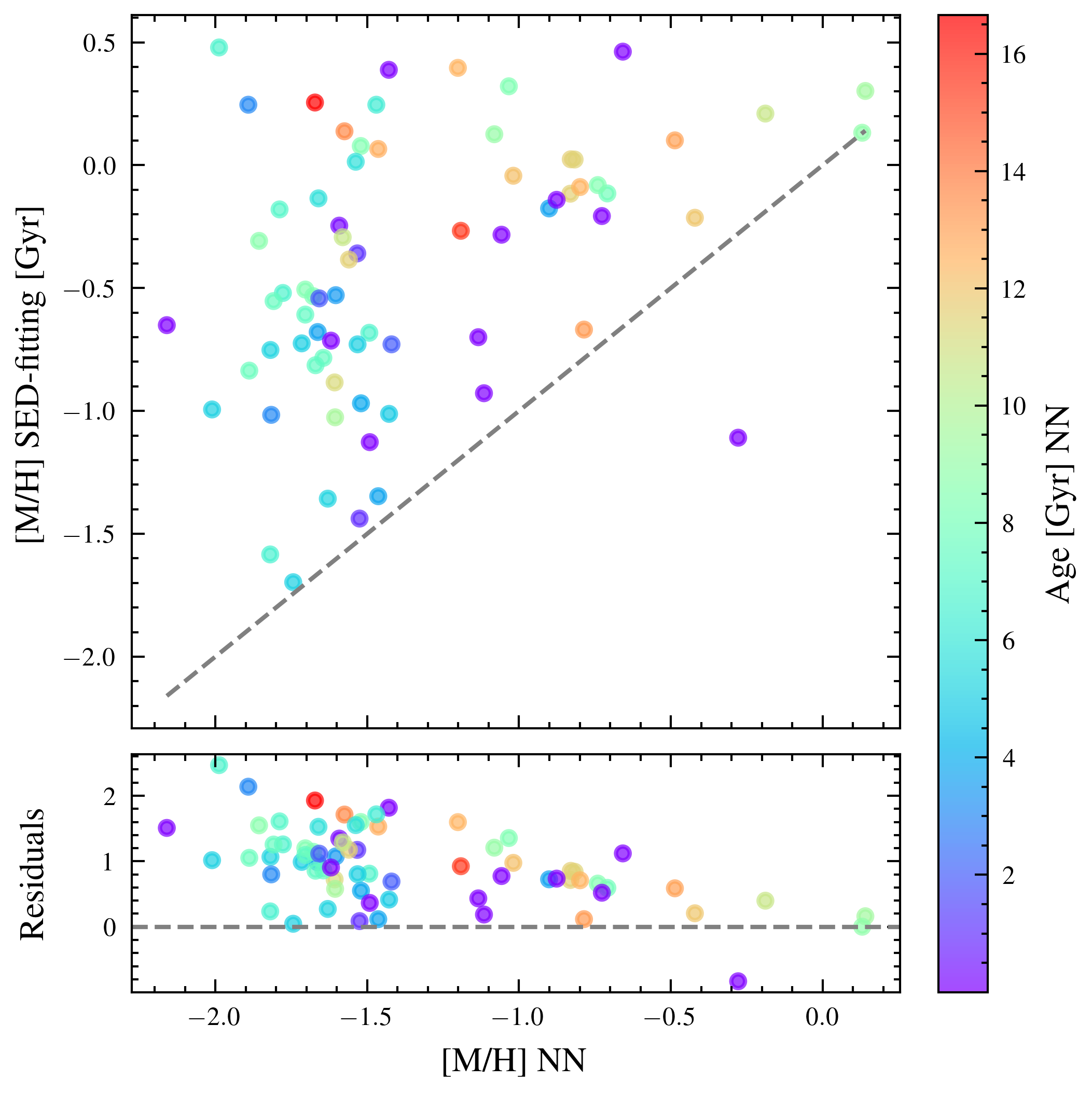}
    \includegraphics[width=0.33\hsize]{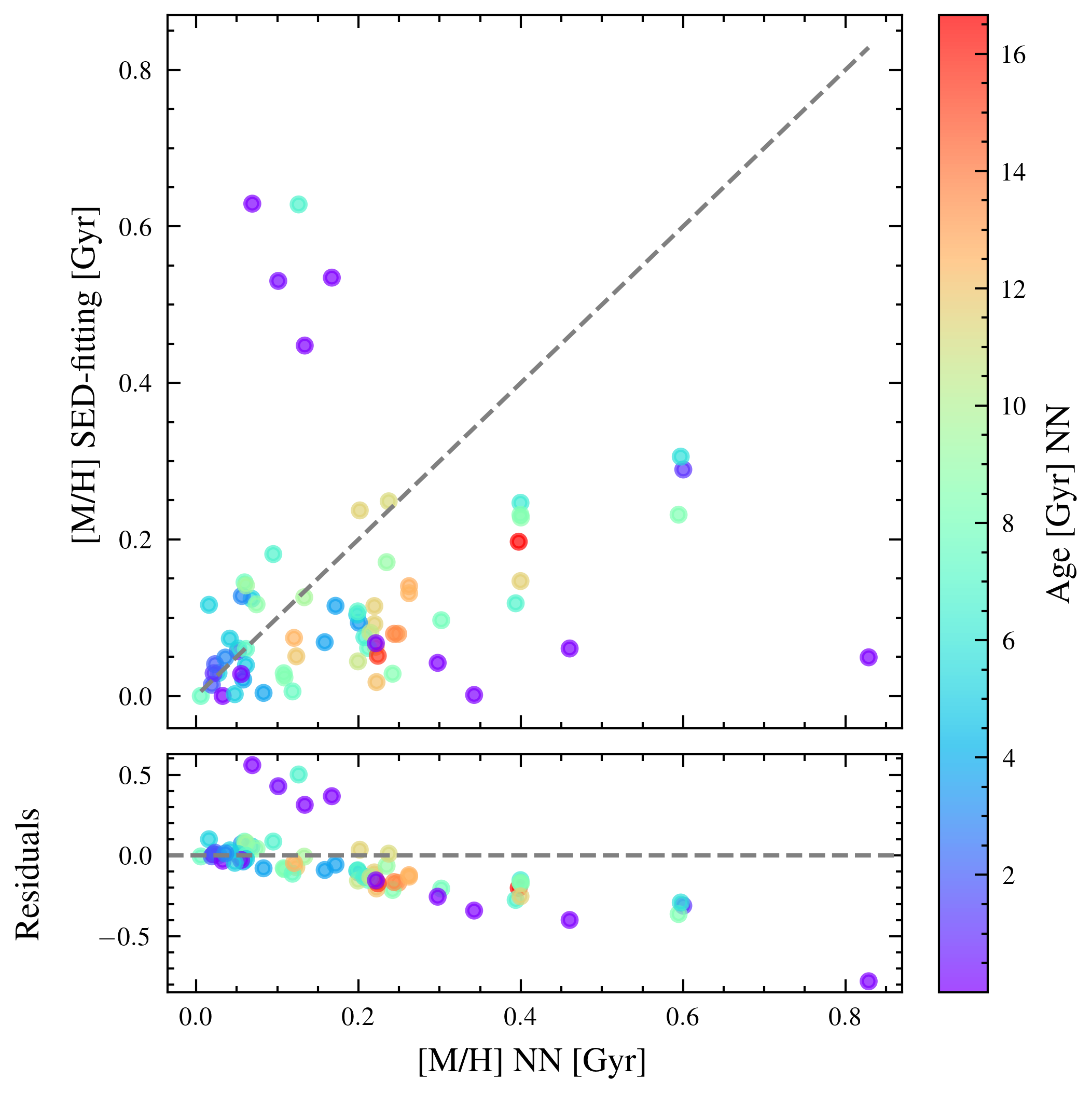}

   \caption{Age (left, color-coded by metallicity), metallicity (middle, color-coded by age) and dust attenuation (right, color-coded by ag) NN predictions versus MUFFIT SED-fitting predictions from Díaz-García (in prep.) for a sample of observed J-PAS galaxies.}
              \label{JPAS}%
    \end{figure*}

Fig. \ref{JPAS} presents the NN predicted ages, metallicities, and dust attenuation compared to the MUFFIT SED-fitting results. The NN predictions yield reasonable values, with ages spanning from very young to very old stellar populations (although a few of them have nonphysically old ages > 12 Gyr), sub-solar metallicities, and intermediate E(B-V) values peaking around 0.2 mag. Compared to the SED-fitting results, significant differences arise: the SED-fitting age predictions are systematically younger, clustered around $\sim$ 2 Gyr; the  metallicity estimates are systematically higher; the E(B-V) values exhibit a more similar distribution across both methods, although the one-to-one-agreement is far from excellent. We recall the reader that the NN is trained to predict SSP while the SED-fitting from Díaz-García et al. (in prep.) uses two-burst CSP, so the discrepancy between the two approaches is expected.

  Interestingly, the NN predicts notably low metallicity values for the observed galaxies, which contrasts with the bias direction seen in the test sample (see Fig.\ref{Met-Ext}). The location of many of the J-PAS  galaxies does fall in the low metallicity and old age region of the UMAP parameter space (see Fig. \ref{UMAP-prop}). One of the possible explanation for such low metallicity estimates could be related to the presence of emission lines, which are not included in the training sample. Indeed, $\sim$ 25\% of the galaxies with NN predicted [M/H] < -1.5 have emission lines with S/N > 3 (Fernández-Ontiveros et al. in prep.). An alternative explanation involves more complex star formation histories, where recent star formation leads to bluer colors. This could influence the metallicity estimates and it is, in fact, consistent with the predictions for CSP outlined in Appendix  \ref{Appendix:CSP}. In addition, subtle calibration differences  between the synthetic and observed photometry could also impact the NN estimates (see \citealt{Boris2007}). This result underscores the importance of carefully considering the limitations of NN models trained on data that differ from the target observations, warning against their blind application. We leave for future work the inclusion of more complex SFHs or emission lines in the training sample, due to the complexity of properly modeling the nebular component.

\section{Conclusions}
\label{sect:conclusions}

In this work we present a NN to estimate SP properties (age, metallicity and dust attenuation) of galaxies from J-PAS photometry. The NN is trained with synthetic photometry from SSP models from three state-of-the-art popular libraries (E-MILES, XSL, CB19). We add realistic noise to the synthetic photometry by applying gaussian variations with an amplitude of typical errors in the mini-JPAS data.  Our main conclusions are:

\begin{itemize}
\item The properties estimated by the NN on the test sample are accurate but very dependent on the S/N of the synthetic photometry. For $i$ $\sim$ 17 mag we obtain  median bias $\mu$= (0.01, 0.00, 0.00), scatter $\sigma_{NMAD}=$(0.23 dex, 0.29 dex, 0.04 mag), percentage of outliers $f_{o}$= (17\%, 24\%, 1\%) and pearson coefficient $P$ = (98, 95 , 98), for $\log t$ [yr], [M/H] and E(B-V), respectively (see Fig. \ref{Metrics}).
\item The NN estimates on the test sample (limited to galaxy spectra built from the CB19 library) outperform the SED-fitting for all the parameters and metrics (see Fig. \ref{Metrics}), demonstrating the great potential of NN for SSP derivation.
\item One of the advantages of this approach is that age, metallicity and dust attenuation are derived independently and, indeed, we observe no evident correlations in the prediction errors on the age-dust-metallicity planes (see Fig. \ref{degeneracies}).
\item We find no clear dependence on the performance of the NN when applied to different SSP libraries (see Fig. \ref{SSP-model}), except for very low ages and metallicities.
\item Combining the three SSP libraries provides better results than training the NN with a single library. It also improves the overlap  between real J-PAS observations and synthetic photometry in the 2D UMAP parameters space (see Fig. \ref{UMAP})
\item When applied to J-PAS photometry of real galaxies, the NN predicted metallicities are significantly lower than  those estimated via SED-fitting using two-burst CSP (see Fig. \ref{JPAS}, while ages are systematically younger).

\item Possible reasons for this discrepancy are discussed in Sect.  \ref{sec:caveats} and include the presence of emission lines (affecting approximately $\sim$25\% of the galaxies with predicted [M/H] < 1–1.5), calibration differences between the synthetic and observed photometry, the effect of redshift, or the existence of more complex SFHs than those represented in the training set. Several of these obstacles would be mitigated in the analysis of globular clusters.
\end{itemize}

Despite these challenges, this pilot study demonstrates the strong potential of NN for estimating SSP parameters from photometric data. The method performs reliably across a wide magnitude range and offers a flexible foundation that can be extended to more complex and realistic scenarios. Future improvements — such as incorporating emission lines, more diverse SFHs, and observed-frame effects — will further enhance its applicability to current and upcoming large photometric surveys.

\begin{acknowledgements}

HDS acknowledges financial support by RyC2022-030469-I grant, funded by MCIU/AEI/10.13039/501100011033  and FSE+ and the Spanish Ministry of Science and Innovation and the European Union - NextGenerationEU through the Recovery and Resilience Facility project ICTS-MRR-2021-03-CEFCA and financial support provided by the Governments of Spain and Arag\'on through their general budgets and the Fondo de Inversiones de Teruel.     
PC acknowledges support from 
Conselho Nacional de Desenvolvimento Cient\'ifico e Tecnol\'ogico (CNPq) under grant 310555/2021-3 and from Funda\c{c}\~{a}o de Amparo \`{a} Pesquisa do Estado de S\~{a}o Paulo (FAPESP) process number 2021/08813-7.

This research has made use of the SVO Filter Profile Service "Carlos Rodrigo", funded by MCIN/AEI/10.13039/501100011033/ through grant PID2023-146210NB-I00. The authors gratefully acknowledge the computer resources at Artemisa, funded by the European Union ERDF and Comunitat Valenciana as well as the technical support provided by the Instituto de Física Corpuscular, IFIC (CSIC-UV).

L.A.D.G. acknowledges financial support from the State Agency for Research of the Spanish MCIU through 'Center of Excellence Severo Ochoa' award to the Instituto
de Astrofísica de Andalucía (CEX2021-001131-S) funded by MCIN/AEI/10.13039/501100011033 and to PID2022-141755NB-I00. J.M.V. acknowledges financial support from the Spanish AEI grant PID2022-136598NB-C32. I.B. has received funding from the European Union's Horizon 2020 research and innovation programme under the Marie Sklodowska-Curie Grant agreement ID n.º 101059532. SGL acknowledges the financial support from the MICIU with funding from the European Union NextGenerationEU and Generalitat Valenciana in the call Programa de Planes Complementarios de I+D+i (PRTR 2022) Project (VAL-JPAS), reference ASFAE/2022/025.
This work is part of the research Project PID2023-149420NB-I00 funded by MICIU/AEI/10.13039/501100011033 and by ERDF/EU.
This work is also supported by the project of excellence PROMETEO CIPROM/2023/21 of the Conselleria de Educación, Universidades y Empleo (Generalitat Valenciana). AAC acknowledges financial support from the Severo Ochoa grant CEX2021- 001131-S funded by MCIN/AEI/10.13039/501100011033 and from the project PID2023-153123NB-I00, funded by MCIN/AEI. The work of V.M.P. is supported by NOIRLab, which is managed by the Association of Universities for Research in Astronomy (AURA) under a cooperative agreement with the U.S. National Science Foundation. GB acknowledges financial support from Universidad Nacional Aut\'onoma de M\'exico through grants DGAPA/PAPIIT IG100319, BG100622 and IN106124.
Partially based on observations made with the JST250 telescope and JPCam at the Observatorio Astrof\'{\i}sico de Javalambre (OAJ), in Teruel, owned, managed, and operated by the Centro de Estudios de F\'{\i}sica del Cosmos de Arag\'on (CEFCA). We acknowledge the OAJ Data Processing and Archiving Department (DPAD) for reducing and calibrating the OAJ data used in this work.

Funding for the J-PAS Project has been provided by the Governments of Spain and Arag\'on through the Fondo de Inversiones de Teruel; the Aragonese Government through the Research Groups E96, E103, E16\_17R, E16\_20R, and E16\_23R; the Spanish Ministry of Science and Innovation (MCIN/AEI/10.13039/501100011033 y FEDER, Una manera de hacer Europa) with grants PID2021-124918NB-C41, PID2021-124918NB-C42, PID2021-124918NA-C43, and PID2021-124918NB-C44; the Spanish Ministry of Science, Innovation and Universities (MCIU/AEI/FEDER, UE) with grants PGC2018-097585-B-C21 and PGC2018-097585-B-C22; the Spanish Ministry of Economy and Competitiveness (MINECO) under AYA2015-66211-C2-1-P, AYA2015-66211-C2-2, and AYA2012-30789; and European FEDER funding (FCDD10-4E-867, FCDD13-4E-2685). The Brazilian agencies FINEP, FAPESP, FAPERJ and the National Observatory of Brazil have also contributed to this project. Additional funding was provided by the Tartu Observatory and by the J-PAS Chinese Astronomical Consortium.
\end{acknowledgements}

\bibliographystyle{aa}
\bibliography{sample}

\begin{appendix}

\section{Optimization of NN hyperparameters}

\label{Appendix:models}
\begin{figure*}
   \centering
   \includegraphics[width=\hsize]{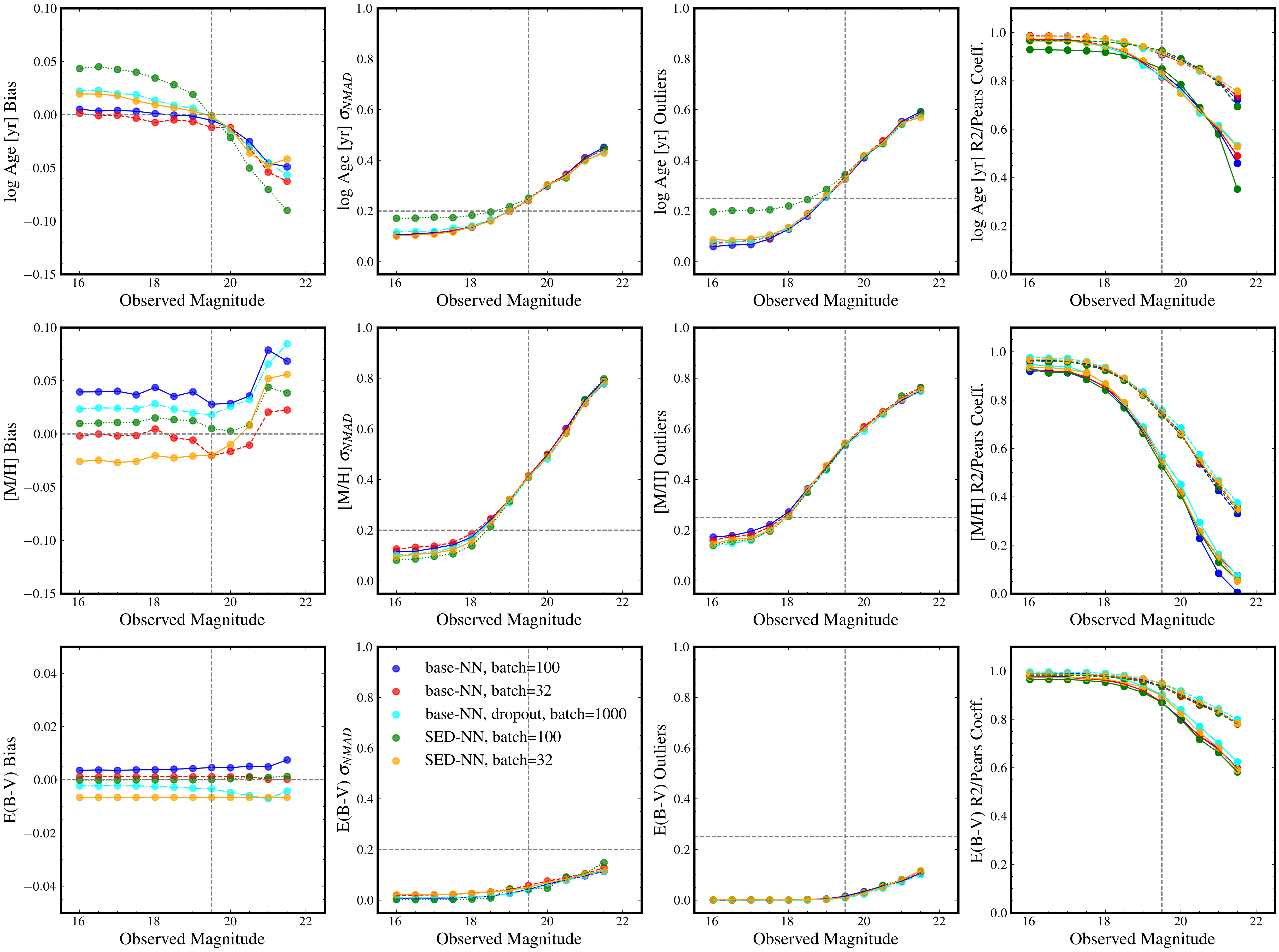}
   \caption{Comparison of the metrics (bias, scatter, fraction of outliers and Pearson (dashed line) and R2 (thick line) coefficients - from left to right) for different parameters (age, metallicity and E(B-V) - top to bottom) as a function of galaxy $i$-band magnitude, obtained for different training strategies, as stated in the legend (details in Sect. \ref{sect:NN}).}
   \label{A-Metrics}%
    \end{figure*}

Neural network hyperparameters can have a significant impact on model performance; however, exploring the full hyperparameter space is computationally expensive and often impractical. For this reason, we limited our investigation to a representative set of configurations detailed in section \ref{sect:NN}. These included variations in the NN architecture (base-NN or SED-NN), $batch~size$ (32, 100, 1000) and the inclusion of $dropout$.  The performance of each setup, reported as the metrics obtained for the predicted stellar population parameters as a function of magnitude (similar to Fig.\ref{Metrics}), is summarized in Fig \ref{A-Metrics}. The main differences are seen in the bias values, while the fraction of outliers or Pearson/R2 coefficients are less affected by the hyperparameter choice. Throughout the paper, we show the results obtained with the base-NN model with a $batch~size =32$ (red line).

\section{Loss curves}
\label{Appendix:Loss}

\begin{figure*}
   \centering
   \includegraphics[width=0.33\hsize]{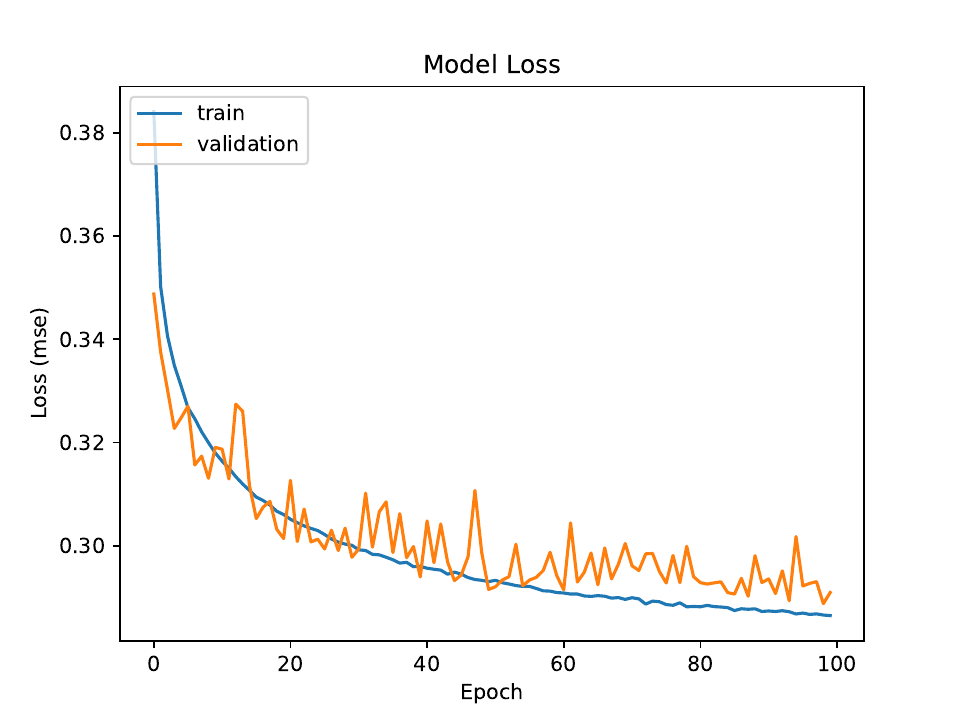}
  \includegraphics[width=0.33\hsize]{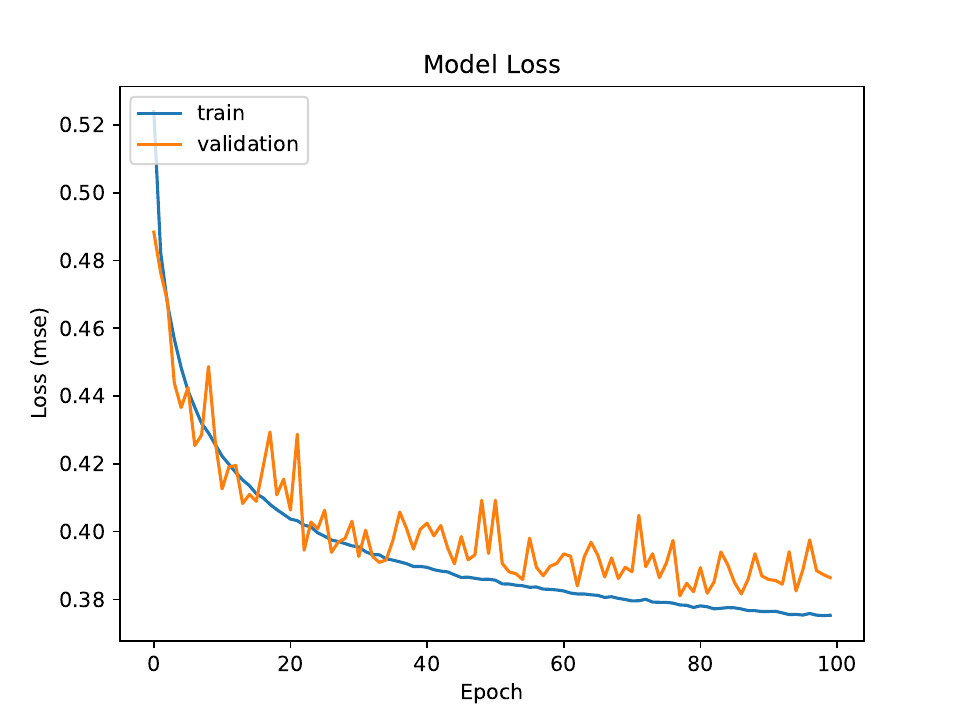}
   \includegraphics[width=0.33\hsize]{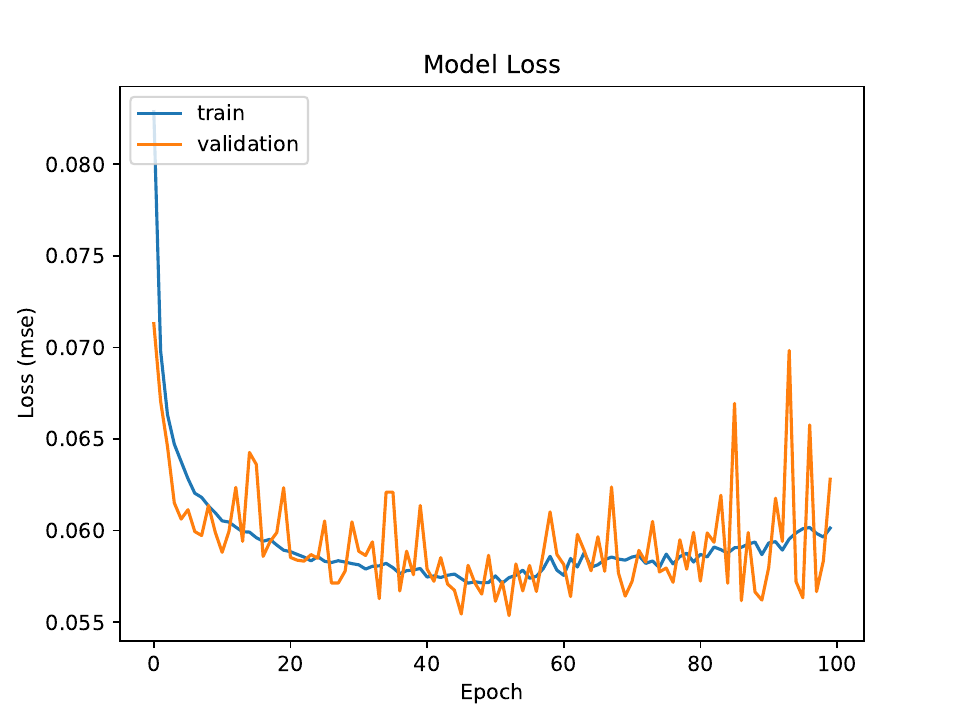}  
   \caption{Loss curves (loss function as a function of training epoch) for the train (blue) and validation (orange) samples for the age (left), metallicity (middle) and dust attenuation (right) NN models. }
   \label{A-Loss}%
    \end{figure*}

Fig \ref{A-Loss} reports the loss curves for the optimal models to illustrate the training dynamics, showing very similar behavior between the training and validation sets. In particular, the E(B–V) model displays an excellent agreement, while the age and [M/H] models show a mild tendency towards overfitting, with the training loss slightly lower than the validation loss. However, the differences remain small ($\le$ 0.1 in MSE) and do not significantly affect the model performance (we have checked and the differences between the predictions of the `best' - the model that minimizes the loss in the validation sample - and the final models are negligible).

\section{Test on Composite stellar populations}
\label{Appendix:CSP}

The NN presented in this work are trained with SSP models, which are a simplistic approximation of the star formation histories of real galaxies. To assess the impact of our estimates on more complex SFHs, we use a set of Composite Stellar Population (CSP) models consisting on:

\setlength{\parskip}{0pt}

\renewcommand{\labelenumi}{\alph{enumi})}  

\begin{enumerate}
\item Three 1\,Gyr burst with different metallicities (Z=0.04, 0.08, 0.17)
\item An initial 1\,Gyr burst at t=0 with metallicity Z=0.008 followed by a second 1\,Gyr burst at different ages  (1, 4, 7, 10 Gyr) and with Z=0.017.
\item  A constant SFR with metalicity Z=0.017
\end{enumerate} 

\begin{figure*}[htbp]
    \centering
 \includegraphics[width=0.9\hsize]{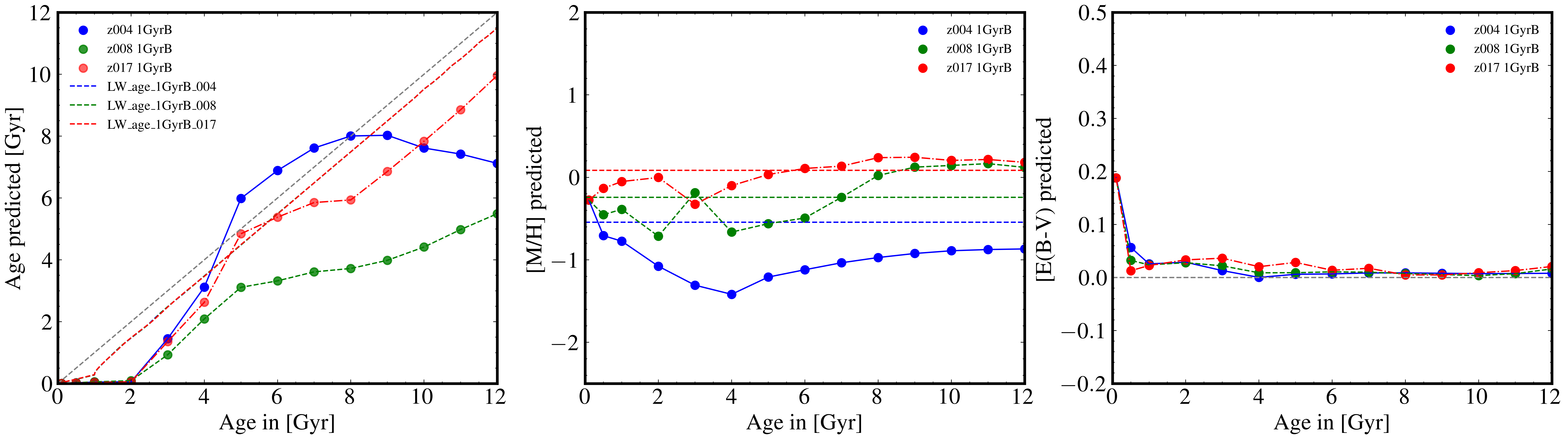}
  \includegraphics[width=0.9\hsize]{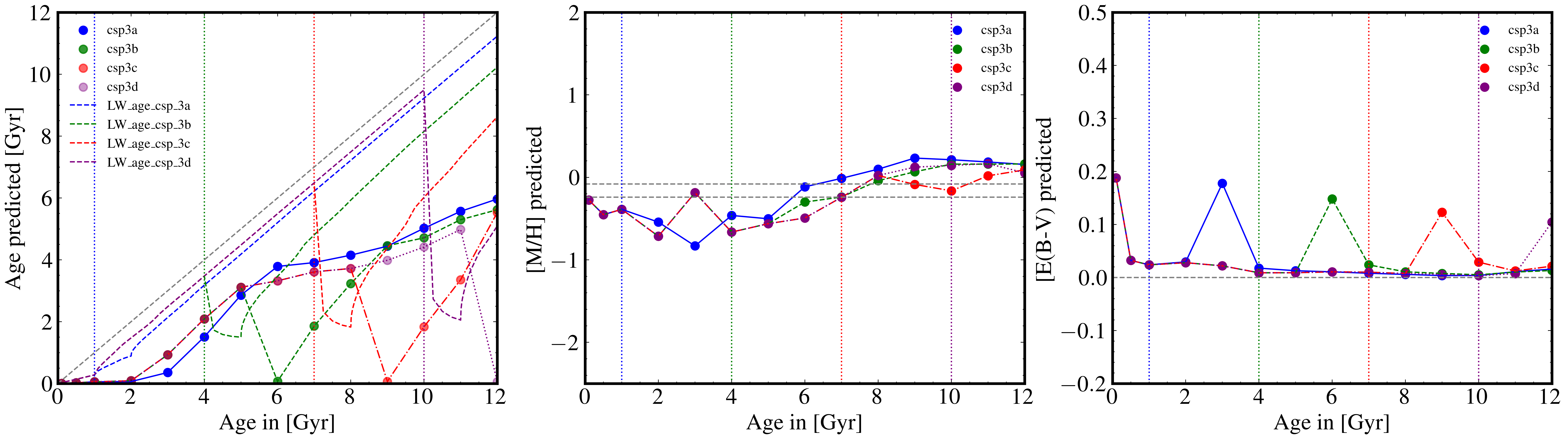}
   \includegraphics[width=0.9\hsize]{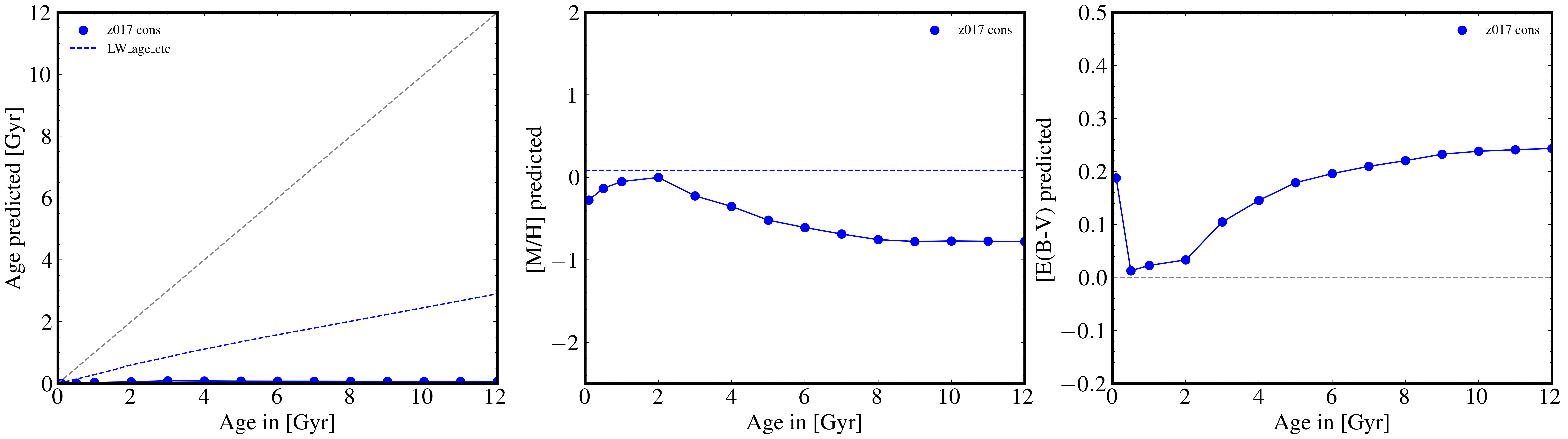}
    \caption{NN predictions for different CSP models -from top to bottom: 1 Gyr bursts, two 1 Gyr bursts, constant SFR (colors correspond to different flavours, as stated in the legend). Age, metallicity and dust attenuation predictions are shown (left to right). The dashed lines show the light-weighted ages, metallicities or E(B-V). The vertical dashed lines in the middle panel show the time when the second burst takes place.  }
    \label{CSP}
\end{figure*}

No dust attenuation is applied to  avoid further degeneracies. Fig. \ref{CSP} shows the NN predictions for these CSP. The predictions for the 1\,Gyr burts (upper panel) show a significant dependence on the metallicity, being more robust for the highest metallicity case (Z=0.017, red line). For the other two metallicities, the ages are usually underpredicted (specially for Z=0.008), while the metallicities  show average deviations of $\sim$ 0.5 dex (up to 1dex for the Z=0.008 case). On the contrary, the dust attenuation is well constrained (E(B-V)$\sim$0) in all cases. The results for the two-burst CSP are a bit more complex (middle panel). In general, ages are underpredicted, but metallicities are constrained within $\sim$ 0.5 dex. Interestingly, both the age and E(B-V) predictions show clear fluctuations $\sim$ 2 Gyr after the second burst takes place (marked as dashed vertical line). Finally, for the constant SFR case (lower panel), none of the parameters is well constrained, specially after 2Gyr. The metallicity is underpredicted ($\gtrsim$ 0.5), the dust attenuation overpredicted (up to 0.2 mag difference) and the predicted age is always $\sim$ 0. Since the NN was trained on SSP, the challenges encountered when recovering the properties of CSPs are not unexpected. Rather than a limitation, this points to a clear direction for future work: incorporating more complex and realistic SFHs into the train and test datasets to improve the model's performance on more representative galaxy populations.

\section{SED-fitting plots}
\label{Appendix:SED-fitting}

\begin{figure*}[htbp]
    \centering
 \includegraphics[width=0.32\hsize]{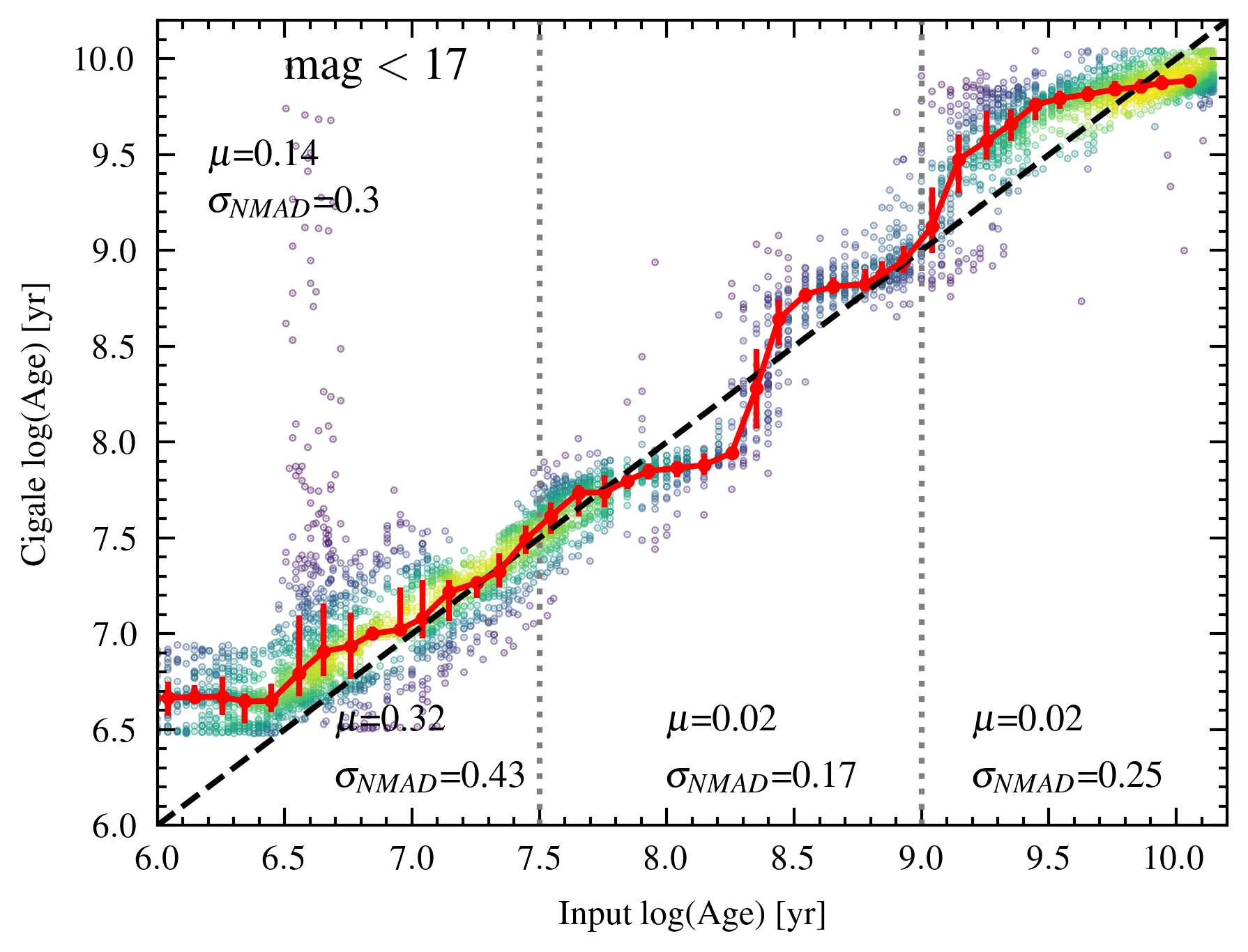}
  \includegraphics[width=0.32\hsize]{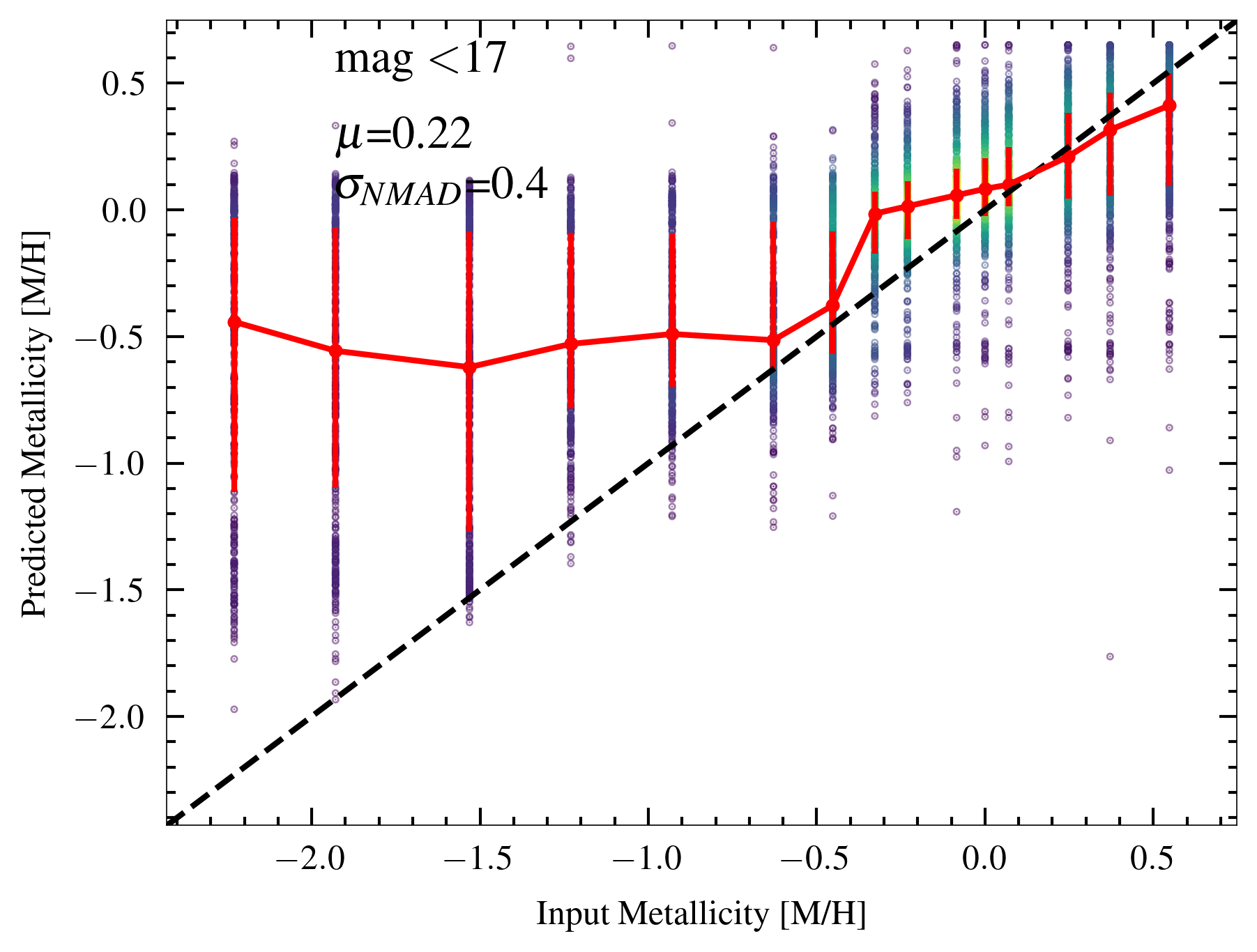}
   \includegraphics[width=0.32\hsize]{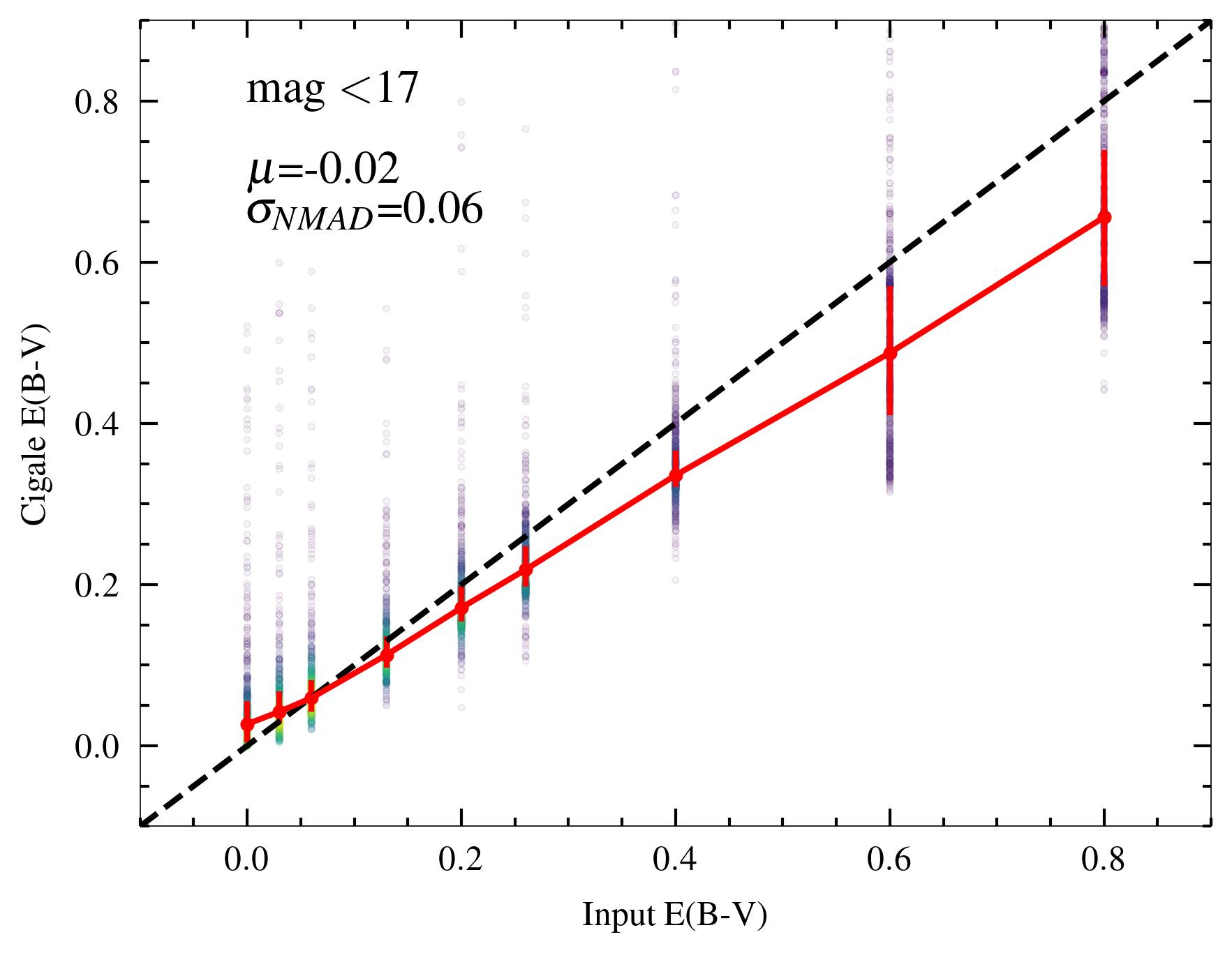}
    \caption{Same as Fig. \ref{Age} and \ref{Met-Ext} but for the output obtained with  CIGALE SED-fitting  (see Sect.\ref{sect:SED-fitting} for details).}
    \label{fig-App:SED-fittig}
\end{figure*}

In figure \ref{Metrics} we compared the metrics summarising the performance of the results obtained by the NN and the SED-fitting with CIGALE on the same test set. To shed more light on the SED-fitting estimates, Fig. \ref{fig-App:SED-fittig} shows the analogue of Figs. \ref{Age} and \ref{Met-Ext} for the CIGALE bayesian outputs. There is a clear quantization of the age predictions, the E(B-V) is slightly underestimated and the [M/H] are clearly overestimated for [M/H] < 0.

\section{UMAP properties}
\label{Appendix:UMAP}

\begin{figure*}[htbp]
    \centering
 \includegraphics[width=0.45\hsize]{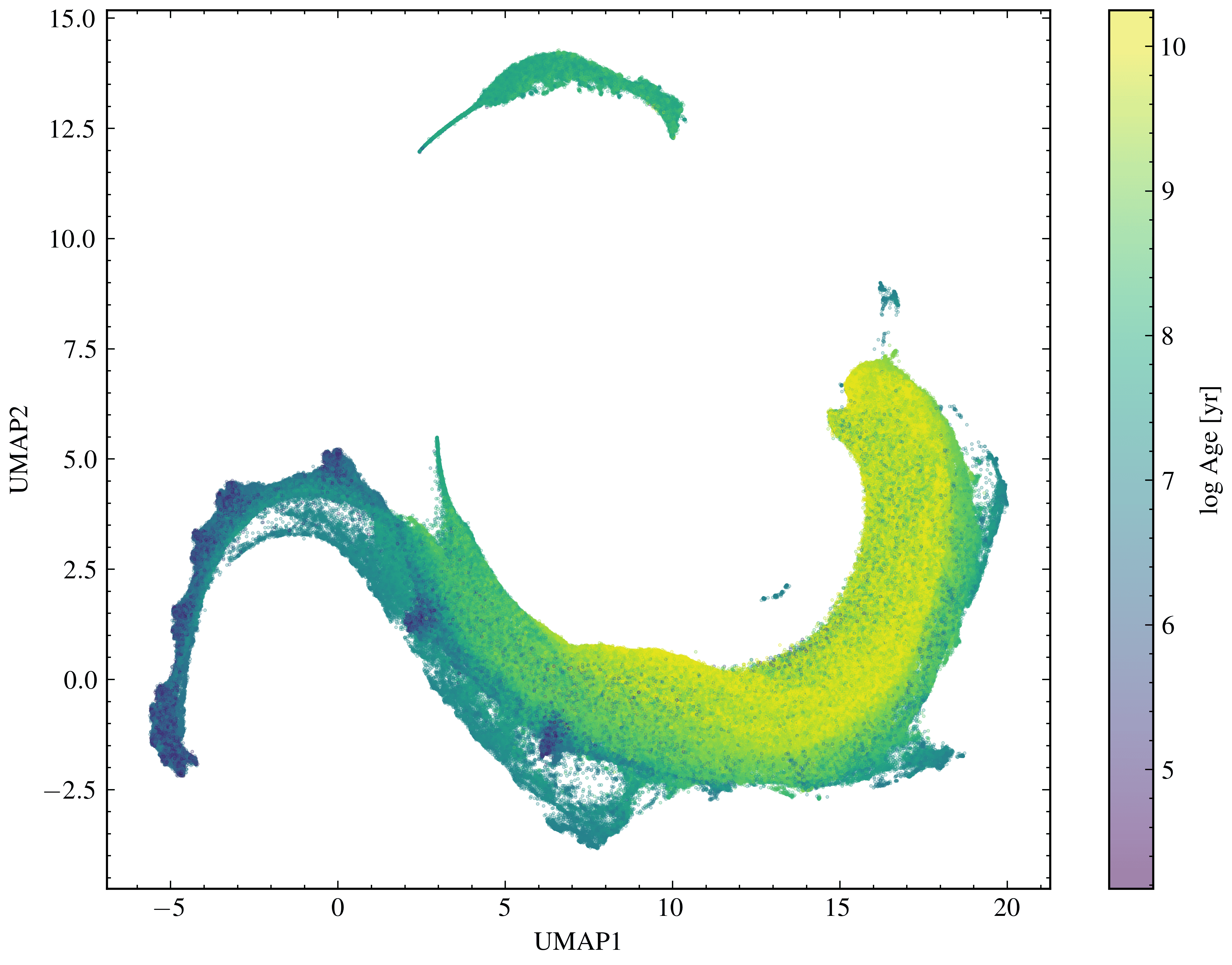}
  \includegraphics[width=0.45\hsize]{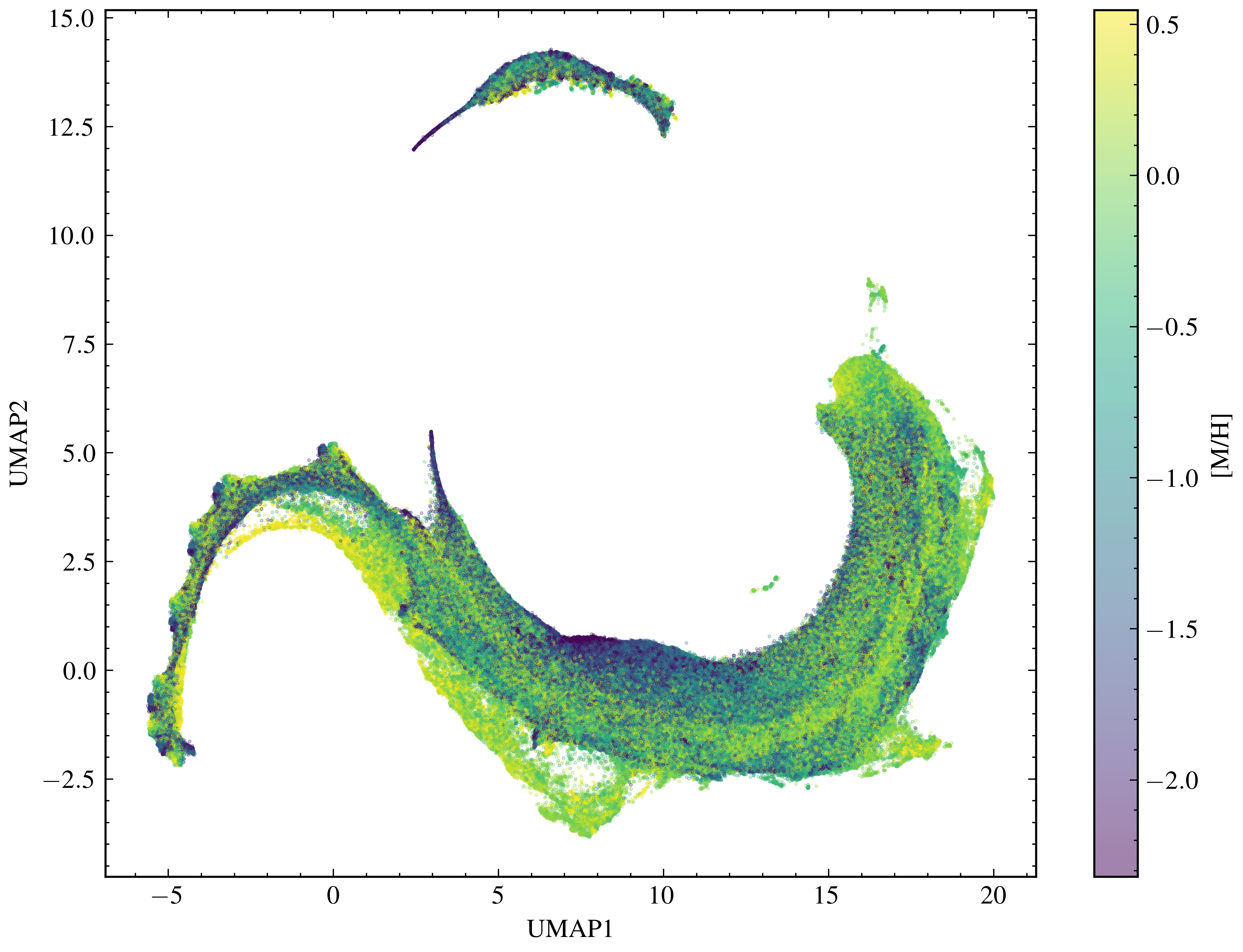}
   \includegraphics[width=0.45\hsize]{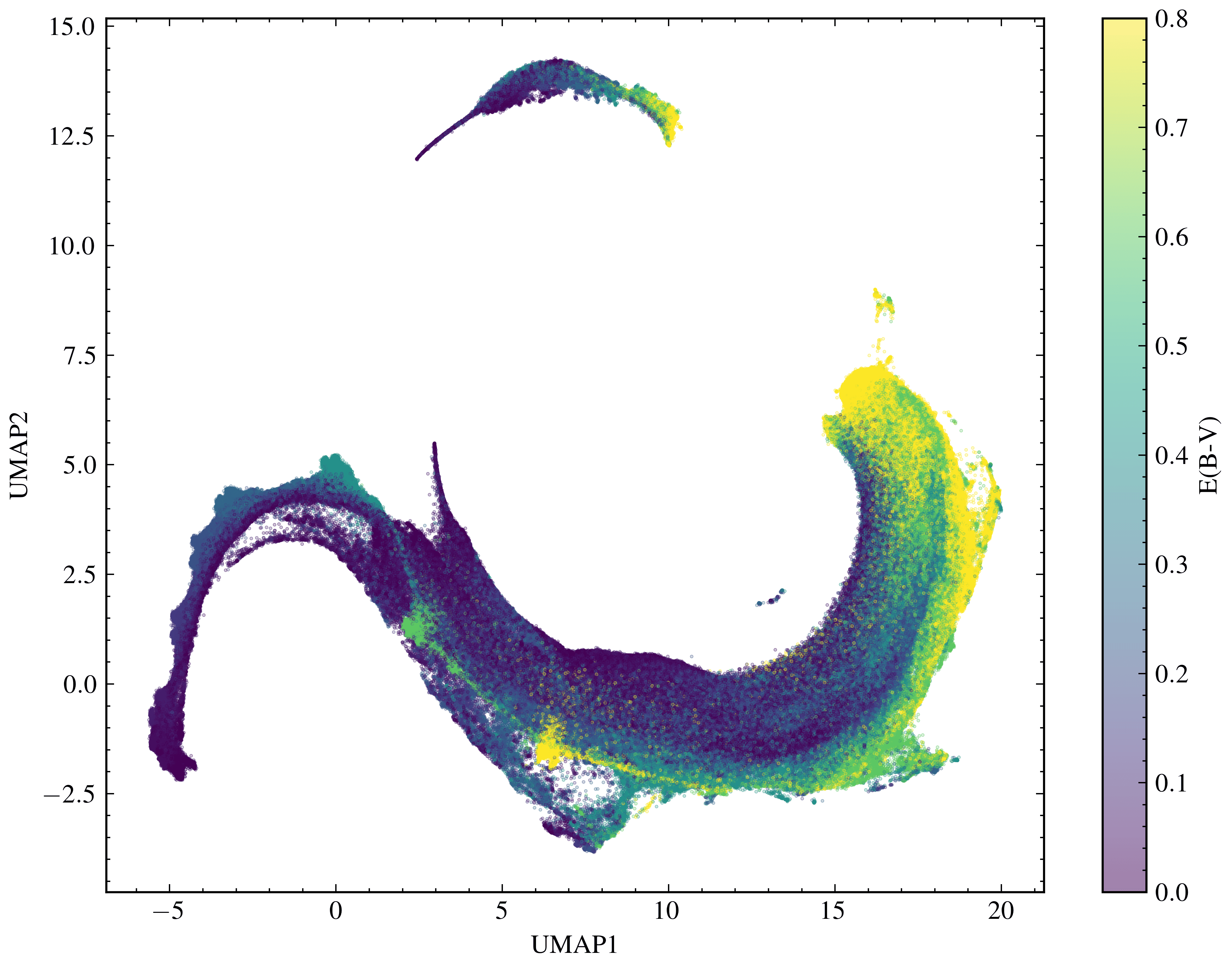}
    \includegraphics[width=0.45\hsize]{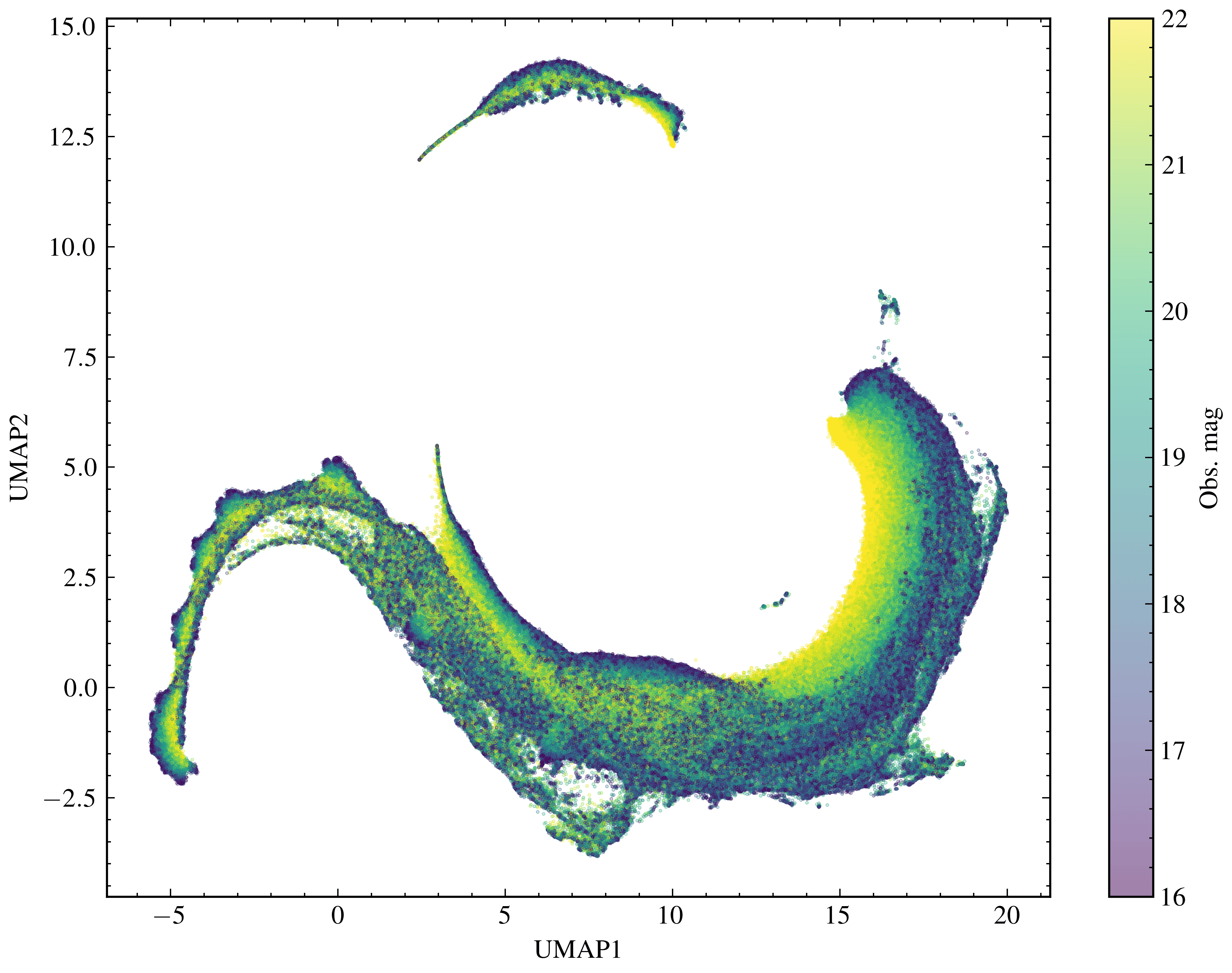}

    \caption{Projection of the synthetic photometry in the 2D parameters space obtained with UMAP dimensionality reduction algorithms, color-coded by age  (upper left), metallicity (upper right), dust attenuation (bottom left) and observed magnitude (bottom right).}
    \label{UMAP-prop}
\end{figure*}

In section \ref{sect:JPAS} the projection of the J-PAS observations in the 2D UMAP parameters space is shown on top of the SSP synthetic photometry. Fig \ref{UMAP-prop} shows the projection of the synthetic photometry (combining the three SSP libaries) on the same 2D parameter space, colour-coded by SSP parameters and observed magnitude. Notably, many of the J-PAS galaxies fall within regions occupied by models with low [M/H], suggesting that, in this 2D parameter space, the observed galaxy colors are consistent with those of low-metallicity SSPs. While this does not guarantee that the galaxies' metallicities are truly so low, it does support the consistency of our methodology with the model's predictions.

\end{appendix}

\end{document}